\documentclass[a4paper,twocolumn,11pt,accepted=2023-10-07]{quantumarticle}
\pdfoutput=1
\usepackage[utf8]{inputenc}
\usepackage[english]{babel}
\usepackage[T1]{fontenc}
\usepackage{amsmath}
\usepackage{hyperref}
\usepackage[numbers]{natbib}

\usepackage{amssymb,amsfonts}
\usepackage{bm}
\usepackage{url}
\usepackage{mathtools}

\newcommand{\deff}{{\, \vcentcolon = \,}}

\begin{document}

\title{A probabilistic view of wave-particle duality for single photons}

\author{Andrea Aiello}
\affiliation{Max Planck Institute for the Science of Light, Staudtstrasse 2, 91058 Erlangen, Germany}
\orcid{0000-0003-1647-0448}
\email{andrea.aiello@mpl.mpg.de}
\homepage{https://mpl.mpg.de/}
\maketitle

\begin{abstract}
One of the most puzzling consequences  of interpreting quantum mechanics in terms of concepts borrowed from classical physics, is the so-called wave-particle duality. Usually, wave-particle duality is illustrated in terms of  complementarity between path distinguishability and fringe visibility in interference experiments. In this work, we instead propose a new type of complementarity, that between the continuous nature of waves and the discrete character of particles. Using the probabilistic methods of quantum field theory, we show that the simultaneous measurement of the wave amplitude and the number of photons in the same beam of light is, under certain circumstances, prohibited by the laws of quantum mechanics. Our results suggest that the concept of ``interferometric duality'' could be eventually replaced by the more general one of ``continuous-discrete duality''.
\end{abstract}

\section{Introduction}

In classical mechanics a physical system is characterized by a set of parameters, called
\emph{degrees of freedom}, which  define its state or configuration at any  time \cite{GoldsteinBook}. Such a set  may be  either  countable (finite or denumerable), or uncountable. The branch of classical mechanics that studies \emph{discrete} systems with a countable set of degrees of freedom, is called \emph{particle mechanics}. Conversely,  \emph{continuous} systems described by an uncountable set of degrees of freedom, are the subjects of \emph{continuum mechanics} \cite{FetterBook}.
In particle mechanics, a  system is described  by a set of functions of \emph{time} $t$, the   so-called generalized coordinates. In contrast, in continuum mechanics a  system is characterized by a set of functions of \emph{spacetime} points $(x,y,z,t)$,  which are the components of scalar, vector or tensor fields. The oscillations of such fields are called \emph{waves}.

Thus, in classical mechanics a physical system is described  either as discrete or continuous (or part discrete and part continuous), and the two descriptions are mutually exclusive\footnote{This does not mean, for example, that we cannot use coordinates to portray some characteristics of a field. Consider, for example, an electromagnetic wave-packet with energy density $U(\mathbf{r},t)$. Such wave-packet is completely described by the electric and magnetic fields. However,  we can introduce the ``energy center of gravity of the field'' $\mathbf{R} = \mathbf{R}(t)$, defined by $\mathbf{R}(t) = \int \mathbf{r} \, U(\mathbf{r},t)\, \mathrm{d} \mathbf{r}/\int  U(\mathbf{r},t) \, \mathrm{d} \mathbf{r}$ \cite{SchwingerMilton}, to picture the mean position and velocity $\mathbf{V} = \mathrm{d} \mathbf{R}/ \mathrm{d} t$ of the wave-packet. However, the coordinates $\mathbf{R}(t)$ are \emph{emergent} quantities that are not necessary for the complete description of the system.}.
This entails that something described by the laws of classical physics, exhibits either wave or particle properties. In quantum mechanics, on the other hand, one and the same system can be described by a set of different physical observables, some of which are wave-like and others particle-like in character.  Hence the celebrated wave-particle duality \cite{GalindoI,PhysRevD.19.473,Grangier_1986,Aspect1987,Aspect1990,Scully1991,OperationalQP,PhysRevX.11.031041,Qian:18,PhysRevResearch.2.012016}.

One characteristic that distinguishes waves from particles is that waves exhibit interference,  particles do not. As a result, most of the works about  wave-particle duality  focus on the  analysis of interference experiments,
in which the complementarity between the distinguishability of a particle's path inside the interferometer and the visibility of interference fringes at its exit, is quantified by means of some cleverly built inequalities.
In recent years, therefore, wave-particle duality has become  synonymous with interferometric duality     \cite{PhysRevA.51.54,PhysRevLett.77.2154,PhysRevA.56.55,Durr1998,PhysRevLett.81.5705,PhysRevA.58.3477,PhysRevA.60.4285,RempeAJP}.

However, there is also another characteristic that distinguishes waves from particles, and that is that waves  have a continuous character, i.e., their amplitudes vary smoothly, whereas particles  have a discrete character, i.e., they can be counted.
This simple consideration enables us to analyse wave-particle duality not from the conventional point of view of complementarity between  path distinguishability and fringe visibility in an interference experiment, but from the new perspective of correlations between continuous and discrete random variables, which represent the values taken by certain wave-like and particle-like observables, respectively, when a measurement is made on one and the same system.
This makes it possible to look at  wave-particle duality in a new and fundamental way.

\begin{figure}[ht!]
  \centering
  \includegraphics[scale=3,clip=false,width=1\columnwidth,trim = 0 0 0 0]{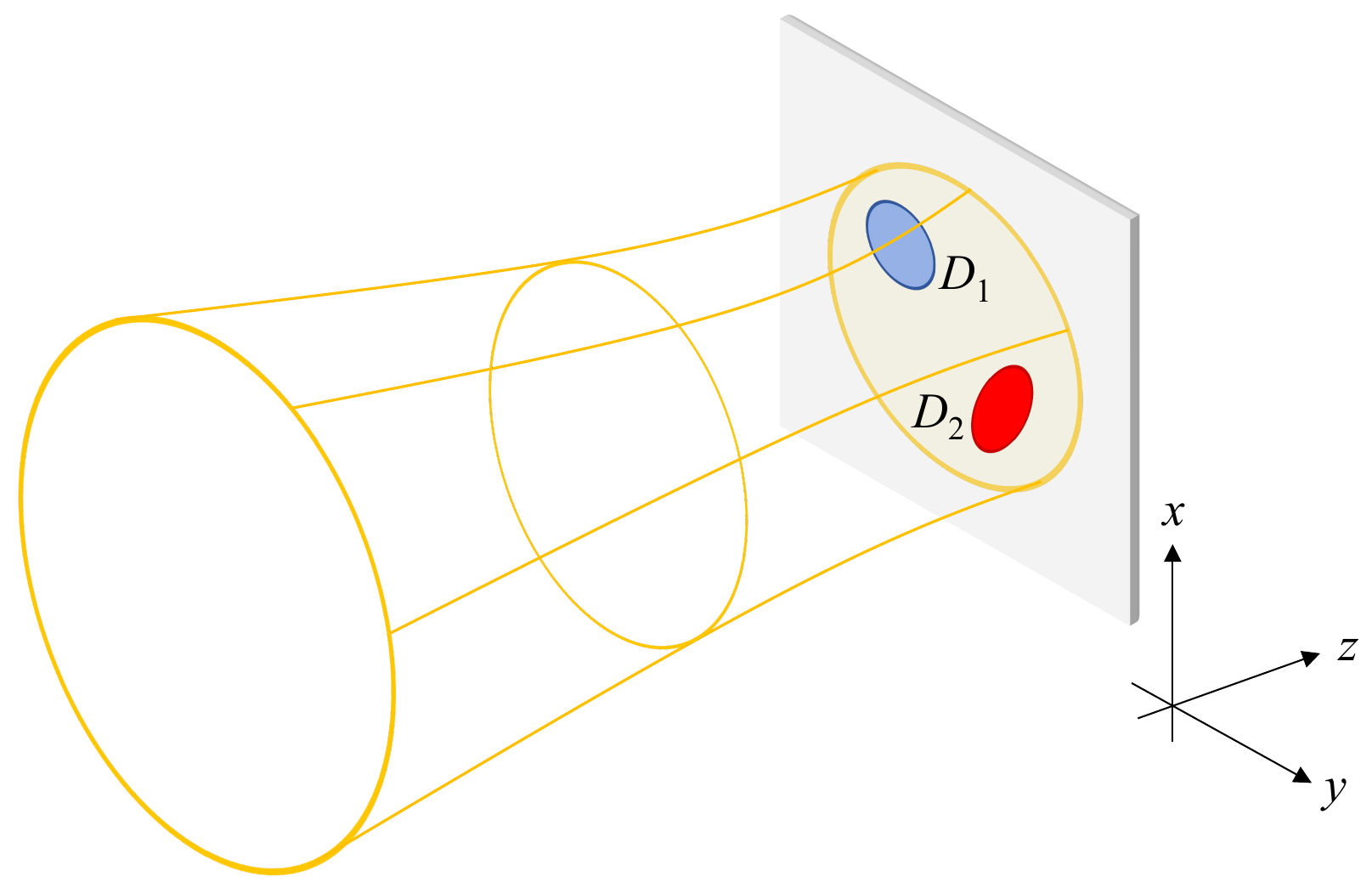}
  \caption{A pictorial representation of the collimated light beam impinging upon the detection screen (gray surface). Blue and red spots on the screen depict the active surfaces of the two detectors $D_1$ and $D_2$. }\label{fig1}
\end{figure}

To be more specific, let us consider the following experiment. A collimated beam of light prepared in a single-photon Fock state \cite{LoudonBook,Scarani1998,vedral2021quantum,Couteau2023},
impinges upon a detection screen, as shown in  Fig. \ref{fig1}.
On this screen there are two spatially separated detectors, say $D_1$ and $D_2$. Each detector  can be set to either  measure the continuous \emph{W}ave amplitude $\mathcal{W}$  of the electric field (for example, by means of homodyne techniques \cite{Fuwa2015,RevModPhys.92.035005}), or to \emph{C}ount the discrete number $\mathcal{C}$ of photons, of the light falling on it. Depending on how we set up these two detectors, we can measure three different pairs of observables, that is $(\mathcal{W}_1,\mathcal{W}_2), ~ (\mathcal{C}_1,\mathcal{C}_2)$, and $(\mathcal{W}_1,\mathcal{C}_2)$. The last pair is particularly interesting because it represents the simultaneous measurement of a wave-like ($\mathcal{W}_1$) and a particle-like ($\mathcal{C}_2$) observable of the system.
Given the single-photon quantum state of the light illuminating both detectors, and given the Hermitian operators  $\hat{W}_1$ and $\hat{C}_2$ describing the observables $\mathcal{W}_1$ and $\mathcal{C}_2$, respectively, we  use
 von Neumann's spectral theorem  \cite{vonNeumann2018,aiello_arXiv.2110.12930}  to calculate the joint probability distribution for the random variables $W_1$ and $C_2$, which represent the values taken by  $\mathcal{W}_1$ and $\mathcal{C}_2$ when a  measurement is actually performed \cite{RevModPhys.92.035005}.
This probability distribution shows that it is possible to measure a nonzero wave amplitude with  detector $D_1$ and simultaneously to count one photon with detector $D_2$, but such wave amplitude will be \emph{entirely} due to the vacuum field fluctuations at detector's location. In contrast, when  detector $D_1$ measures a wave amplitude which is \emph{not entirely} due to the vacuum field fluctuations, then detector $D_2$ will \emph{always} count zero photons.
There is thus a kind of complementarity between the continuous and discrete nature of the electromagnetic field prepared in a single-photon state: The observation of a non-vacuum-yielded wave amplitude and the count of a photon, are  mutually exclusive. This is the first main results of our work.

Furthermore,  we  find that the values taken by  $\mathcal{W}_1$ and $\mathcal{C}_2$ are correlated in a \emph{nonlinear} manner although the operators $\hat{W}_1$ and $\hat{C}_2$ do commute when the detectors $D_1$   and $D_2$  are spatially separated. This is a purely quantum effect due to the non-localizability of the single-photon field which extends over the surfaces of both detectors. In the jargon of probability theory, we can say that the random variables $W_1$ and $C_2$ are linearly uncorrelated but not independent \cite{MurphyBook2022}.
The key tool we use to reveal this nonlinear correlation is the mutual information \cite{CoverThomas}, a statistical measure that finds numerous applications in contemporary physics (see, e.g., \cite{PhysRevLett.119.200502,PhysRevLett.126.200601}, and references therein). This is our second main result.

The rest of this paper is organized as follows. In section \ref{QFT} we quickly present a phenomenological quantum field theory of paraxial beams of light, and we jot down the quantum states of the field. In section \ref{smeared} we build up and characterize the Hermitian quantum operators representing the wave ($\mathcal{W}$) and particle ($\mathcal{C}$) observables of the electromagnetic field. In section \ref{pd} we briefly review probability theory for quantum operators. Next, in section \ref{wp} first we write down and discuss the formulas for the joint probability distributions for the three pairs of random variables $(W_1,W_2), ~ (C_1,C_2)$, and $(W_1,C_2)$ describing the results of the experiment pictured above.
Then, we apply these formulas to the cases of  vacuum  and  single-photon  input states of the electromagnetic field.
In section \ref{discuss}, we discuss the results obtained in the previous section.
 Finally, in section \ref{conc} we briefly summarise our results and draw some conclusions. Four appendices report detailed calculations of the results presented in the main text.

\section{Quantum field theory of light}\label{QFT}

In this section we give a brief overview of the quantum field theory of paraxial light beams. We also define and illustrate the quantum states of the electromagnetic field that will be used later.

\subsection{Paraxial quantum field operators}

Following closely \cite{aiello_arxiv_2022}, we consider   a monochromatic  paraxial  beam of light of frequency $\omega$,  propagating in the $z$ direction and polarized along the $x$ axis of a given Cartesian coordinate system.
In the Coulomb gauge, the electric field operator can be written as  $\hat{\mathbf{E}}(\mathbf{r},t) = \hat{\Phi} (\mathbf{x},z,t) \, \hat{\mathbf{e}}_x$, where $\mathbf{r} = x \hat{\mathbf{e}}_x + y \hat{\mathbf{e}}_y + z \hat{\mathbf{e}}_z \deff \mathbf{x} + z \hat{\mathbf{e}}_z$ is the position vector and, in suitably chosen units,
\begin{align}\label{a70}
\hat{\Phi} (\mathbf{x},z,t) = \frac{1}{\sqrt{2}}  \left[ e^{- i \omega t} \hat{\phi} (\mathbf{x},z)  +  e^{i \omega t} \hat{\phi}^\dagger (\mathbf{x},z)  \right],
\end{align}
with
\begin{align}\label{n10}
\hat{\phi} (\mathbf{x},z) =  \sum_{\mu} \hat{a}_{\mu} u_{\mu} (\mathbf{x},z).
\end{align}
Here and below  $\mathbf{x} = x \bm{e}_x + y \bm{e}_y$ is the transverse position vector,  and the elements $u_{\mu} (\mathbf{x},z)$ of the  countable set of functions $\{ u_{\mu} (\mathbf{x},z)\}$, are the so-called spatial modes of the field labeled by the index $\mu$, which denotes an ordered pair of integer numbers. For example, $\mu = (n,m), \; (n,m = 0,1,2, \ldots)$, for Hermite-Gauss  modes, and $\mu = (\ell,p), \; (\ell = 0, \pm 1, \pm 2, \ldots, \,p = 0,1,2, \ldots)$, for Laguerre-Gauss modes.
By hypothesis, the spatial  modes are solutions of the paraxial wave equation \cite{GoodmanBook}, and form a complete and orthogonal set of basis functions on $\mathbb{R}^2$, i.e.,
\begin{align}\label{n20}
\sum_{\mu} u_{\mu}(\mathbf{x},z)u_{\mu}^*(\mathbf{x}',z) =   \delta \left( \mathbf{x} - \mathbf{x}' \right) ,
\end{align}
with $\delta \left( \mathbf{x} - \mathbf{x}' \right) = \delta \left( x - x' \right) \delta \left( y - y' \right)$,  and
\begin{align}\label{n30}
\bigl( u_{\mu}, u_{\mu'}\bigr) = \delta_{\mu \mu'},
\end{align}
respectively. Here and hereafter we use the suggestive notation
\begin{align}\label{n40}
(f, g) = \int_{\mathbb{R}^2}  f^*(\mathbf{x},z) g(\mathbf{x},z) \, \mathrm{d} \mathbf{x} ,
\end{align}
where $\mathrm{d} \mathbf{x} = \mathrm{d}x \, \mathrm{d}y$.
As usual, the annihilation and creation operators $\hat{a}_{\mu}$ and $\hat{a}^{\dagger}_{\mu}$, respectively,  satisfy the bosonic canonical commutation relations
\begin{align}\label{n15}
\bigl[ \hat{a}_{\mu}, \; \hat{a}^\dagger_{\mu'} \bigr] = \delta_{\mu \mu'}.
\end{align}
Finally, we remark that from \eqref{a70}--\eqref{n40} it follows that   the dimension of  $\hat{\Phi} (\mathbf{x},z,t)$ is the inverse of a length: $[\hat{\Phi}] = L^{-1}$.

\subsection{Quantum states of the electromagnetic field}

Consider a classical paraxial beam of light carrying the  electric field $\mathbf{E}_\text{cl}(\mathbf{r},t) = \Phi(\mathbf{x},z,t) \, \hat{\mathbf{e}}_x$, where
\begin{align}\label{n50}
\Phi (\mathbf{x},z,t) = \frac{1}{\sqrt{2}}  \left[ e^{- i \omega t} \phi (\mathbf{x},z)  +  e^{i \omega t} \phi^* (\mathbf{x},z)  \right].
\end{align}
Here the scalar field $\phi(\mathbf{x},z)$ is a solution of the paraxial wave equation normalized to
\begin{align}\label{n55}
(\phi, \phi) = \int_{\mathbb{R}^2} \left| \phi(\mathbf{x},z) \right|^2 \, \mathrm{d} \mathbf{x} = 1.
\end{align}
By construction, the classical field $\Phi (\mathbf{x},z,t)$ is equal to the expectation value of the quantum field $\hat{\Phi} (\mathbf{x},z,t)$ with respect to the coherent state $|\{ \phi \} \rangle$, i.e., $\Phi (\mathbf{x},z,t) = \langle\{ \phi \} | \hat{\Phi} (\mathbf{x},z,t) | \{ \phi \} \rangle$,  where  $|\{ \phi \} \rangle = \exp \bigl( \hat{a}^\dagger[\phi] - \hat{a}[\phi]\bigr)| 0 \rangle$, $| 0 \rangle$ is the vacuum state of the electromagnetic field defined by $\hat{a}_\mu | 0 \rangle =0$ for all $\mu$,
\begin{align}\label{n60}
\hat{a}^\dagger[\phi] = \bigl( \hat{\phi}, \phi \bigr) =  \sum_\mu \hat{a}^\dagger_\mu \phi_\mu,
\end{align}
with $\phi_\mu = ( u_\mu, \phi)$ \cite{Deutsch1991,aiello_arxiv_2022}, and \eqref{n10} has been used. Note that since both the modes $u_\mu(\mathbf{x},z)$ and the field $\phi(\mathbf{x},z)$ are solutions of the paraxial wave equation, then the coefficients $\phi_\mu$ are independent of $z$. It is not difficult to show that $\bigl[ \hat{a}[\phi],\hat{a}^\dagger[\psi] \bigr]= \bigl( \phi, \psi \bigr)$, for any pair of normalized fields $\phi(\mathbf{x},z)$ and $ \psi(\mathbf{x},z)$.
The field  $\phi(\mathbf{x},z)$ also determines the (improperly called)  wave function of the photon, defined by $\langle 0 | \hat{\Phi}(\mathbf{x},z,t) | 1[\phi] \rangle = e^{- i \omega t} \phi(\mathbf{x},z)/ \sqrt{2} $, where
\begin{align}\label{n70}
| N[\phi] \rangle = \frac{\bigl( \hat{a}^\dagger[\phi] \bigr)^N}{\sqrt{N!}} | 0 \rangle ,
\end{align}
denotes the $N$-photon Fock state with $N= 0,1,2, \ldots$, such that $\hat{N} |N[\phi] \rangle = N|N[\phi] \rangle $,
\begin{align}\label{n75}
\hat{N}= \sum_\mu \hat{a}^\dagger_\mu \hat{a}_\mu ,
\end{align}
and $\langle N[\phi] | M [\psi] \rangle = \bigl( \phi, \psi \bigr)^N \delta_{NM}$ (see Supplemental Material in \cite{aiello_arxiv_2022} for further details).

\section{Wave-like and particle-like operators}\label{smeared}

In this section, we will construct what we call the ``wave operator'' $\hat{W}$ and the ``particle operator'' $\hat{C}$, which represent, respectively, the amplitude $\mathcal{W}$ of the electric field and the number $\mathcal{C}$ of counted photons of some light beam. In our jargon, a wave operator is simply a Hermitian operator with a \emph{continuous spectrum}, while a particle operator is a Hermitian operator with a \emph{discrete spectrum}.
Herein lies the great conceptual difference between classical and quantum mechanics. In the former, the either continuous or discrete character of a physical system is a property of its description that we fix a priori. In the latter, on the other hand, the state of a physical system is always described by a ray in a Hilbert space, and there are certain physical quantities relative to the system, the so-called observables, some of which are discrete and others continuous in character. Hence the wave-particle duality.

\subsubsection{Wave-like operators}

In quantum field theory, a mathematical  object like  $\hat{\Phi} (\mathbf{x},z,t)$ defined by \eqref{a70}, does not really represent a proper observable, because it is not an Hermitian operator in the Hilbert space $\mathcal{H}$ of the physical states  of the electromagnetic field,  but rather an ``operator valued distribution'' over the Euclidean spacetime $\mathbb{R}^2 \times \mathbb{R}$  \cite{Haag1992}. This can be seen, for example, by showing that $\hat{\Phi} (\mathbf{x},z,t)$ does not map the vacuum state $| 0 \rangle \in \mathcal{H}$ into another state in $\mathcal{H}$. To this end, let us define the vector $|\psi \rangle = \hat{\Phi} (\mathbf{x},z,t) | 0 \rangle$. Then, it is not difficult to show that $| \psi \rangle \notin \mathcal{H}$ because it has not a finite norm:
\begin{align}\label{n80}
\langle  \psi | \psi \rangle & =   \lim_{\mathbf{x}' \to \mathbf{x}} \langle  0|\hat{\Phi} (\mathbf{x},z,t) \hat{\Phi} (\mathbf{x}',z,t) |0 \rangle \nonumber \\[6pt]
 & =  \frac{1}{2} \, \lim_{\mathbf{x}' \to \mathbf{x}} \langle  0| \bigl[ \hat{\phi} (\mathbf{x},z,t), \hat{\phi}^\dagger (\mathbf{x}',z,t)\bigr] |0 \rangle \nonumber \\[6pt]
 & =  \frac{1}{2}  \lim_{\mathbf{x}' \to \mathbf{x}} \delta \left( \mathbf{x} - \mathbf{x}' \right) \nonumber \\[6pt]
 & = \infty,
\end{align}
where \eqref{n20} has been used. This means that the quantum fluctuations (variance) of $\hat{\Phi} (\mathbf{x},z,t)$ in the vacuum state blow up for $\mathbf{x}' \to \mathbf{x}$. Thus, to obtain a bona fide Hermitian operator defined on the vectors in $\mathcal{H}$,  we must to smear out $\hat{\Phi} (\mathbf{x},z,t)$ with a real-valued  test function $F(\mathbf{x},t) \in \mathbb{R}$ \cite{Rowe_1979,BladelBook,https://doi.org/10.48550/arxiv.2302.13742}, namely to take
\begin{align}\label{a100}
\hat{\Phi}[F] = \int_{\mathbb{R}^2 \times \mathbb{R}}  F(\mathbf{x},t) \hat{\Phi} (\mathbf{x},z,t) \, \mathrm{d} \mathbf{x} \, \mathrm{d} t.
\end{align}
In the case of free fields, we can  choose $F(\mathbf{x},t) = \delta(t-t_0) f(\mathbf{x})$  \cite{Haag1992,ItzZub} where we normalize the real-valued function $f(\mathbf{x})$ as
\begin{align}\label{a160}
\int_{\mathbb{R}^2}  f(\mathbf{x}) \, \mathrm{d} \mathbf{x}  =1,
\end{align}
and $t_0$ is any time. Without loss of generality, in the remainder we will set $t_0=0$.
Note that normalization condition \eqref{a160} implies that the dimension of both $f(\mathbf{x})$ and $(f,f)$ is $L^{-2}$. Then we define the smeared field operator $\hat{W} \deff \hat{\Phi}[f]$ by
\begin{align}\label{a170}
\hat{W} \deff \hat{\Phi}[f] = & \; \int_{\mathbb{R}^2}  f(\mathbf{x}) \, \hat{\Phi} (\mathbf{x},z,0) \, \mathrm{d} \mathbf{x} \nonumber \\[6pt]
= & \; \sum_\mu \left( \hat{a}_\mu f_\mu^* + \hat{a}_\mu^\dagger f_\mu \right)/\sqrt{2},
\end{align}
where $f_\mu = (u_\mu, f) = |f_\mu| e^{i \theta_\mu}$ \cite{Haag1992,ItzZub}.  For example, $f(\mathbf{x})$ can be the Gaussian function
\begin{align}\label{a172}
f(\mathbf{x}) = \frac{1}{(a \sqrt{\pi})^2} \, e^{-(x^2 + y^2)/{a^2}},
\end{align}
where $a>0$ is some length. In this case  $\hat{W}$ is  a smoothed form of the field averaged over a region of area $a^2$ \cite{Coleman2019}.

More generally, the physical meaning of $\hat{W}$ is that of a quadrature operator of the electric field, which can be measured by a homodyne detector  \cite{Schleich}. To show this, first we write $\hat{W}$ as
$\hat{W} = \bigl( f, \hat{\Phi}\bigr)$, and then we use the definition \eqref{n60} to obtain
\begin{align}\label{a170bis}
\hat{W} = \frac{1}{\sqrt{2}} \left( \hat{a}[f] + \hat{a}^\dagger[f] \right).
\end{align}
This is indeed the expression of the quadrature Hermitian operator of a single mode of  the electromagnetic field \cite{Schleich,Barnett}.
We remark that since the quadrature operator  has a continuum of eigenvalues\footnote{See, for example, Eq. (11.8) in Ref. \cite{Schleich}.}, then the smeared field $\hat{W}$ has a \emph{continuum spectrum} too.

If one prefers to work with the original operators $\hat{a}_\mu$ associated with the modes $u_\mu(\mathbf{x},z)$, then $\hat{W}$ is given by a weighted sum of quadrature operators. This can be seen by rewriting
\begin{align}\label{quad90}
\hat{W} = \sum_\mu |f_\mu| \, \hat{x}_{\mu},
\end{align}
where, by definition,
\begin{align}\label{quad100}
\hat{x}_{\mu} \deff \bigl( \hat{a}_\mu e^{-i \theta_\mu} + \hat{a}_\mu^\dagger e^{i \theta_\mu} \bigr) /\sqrt{2},
\end{align}
is  the quadrature Hermitian operator of the field component with respect to the mode $u_\mu(
\mathbf{x},z)$ \cite{RevModPhys.92.035005}.

From \eqref{a160} and \eqref{a170} it follows that the dimension of  $\hat{W} = \hat{\Phi}[f]$ is $L$, as that of the original operator $\hat{\Phi} (\mathbf{x},z,0)$. For later purposes, it is useful to calculate the \emph{finite} variance $\sigma^2$ of $\hat{W}$ with respect to the vacuum state:
\begin{align}\label{n105}
\sigma^2 & = \langle 0|\hat{W}^2|0 \rangle \nonumber \\[6pt]
& =  \left(f,f \right)/2.
\end{align}

In the remainder we will consider a set of $M$ different test functions $\{ f_n(\mathbf{x}) \} = \{f_1(\mathbf{x}), \ldots, f_M(\mathbf{x}) \}$, each normalized according to \eqref{a160}. The function $f_n(\mathbf{x})$ characterizes the action of detector $D_n$, when the latter is set to measure the  amplitude of the electric field of light falling on it  (see, e.g., \cite{Rosenfeld_1933,PhysRev.78.794} and \S 9 of \cite{Heitler} for a thorough discussion about measurements of the strength of a quantum field).

In practice, each detector $D_n$ has a limited active surface area. Let $\mathcal{D}_n \in \mathbb{R}^2, ~ (n=1,2,\ldots, M)$ denote the domain in the $xy$-plane occupied by the active surface of detector $D_n$. In principle,  $\mathcal{D}_n$ may have any shape, we only require that $\mathcal{D}_n \cap \mathcal{D}_m = \emptyset$ if $m \neq n$.
With $\bm{1}_{n}(\mathbf{x})$ we denote the indicator function of the domain $\mathcal{D}_n$  defined by
\begin{align}\label{x10}
\bm{1}_{n}(\mathbf{x}) = \left\{
                       \begin{array}{ll}
                         1, & \hbox{for $\mathbf{x} \in \mathcal{D}_n$,} \\[6pt]
                         0, & \hbox{for $\mathbf{x} \not\in  \mathcal{D}_n$.}
                       \end{array}
                     \right.
\end{align}
Note that, by definition,
\begin{align}\label{x20}
\bm{1}_{n}(\mathbf{x}) \bm{1}_{m}(\mathbf{x}) =  \delta_{nm} \bm{1}_{n}(\mathbf{x}).
\end{align}
Then, we can take $f_n(\mathbf{x}) = \bm{1}_{n}(\mathbf{x}) f(\mathbf{x} - \mathbf{x}_n)$, where $f(\mathbf{x} - \mathbf{x}_0)$ is any smooth function concentrated in the neighborhood of $\mathbf{x}_0$, such that
\begin{align}\label{supp}
\int_{\mathbb{R}^2}  f_n(\mathbf{x} ) \, \mathrm{d} \mathbf{x} & =   \int_{\mathcal{D}_n}  f(\mathbf{x} - \mathbf{x}_n) \, \mathrm{d} \mathbf{x} \nonumber \\[6pt]
& \approx  \int_{\mathbb{R}^2}  f(\mathbf{x} - \mathbf{x}_n) \, \mathrm{d} \mathbf{x}  = 1 .
\end{align}
With this choice we have $f_n(\mathbf{x}) f_{m}(\mathbf{x}) = 0$ for $n \neq m$.

Using \eqref{a170} with $f = f_n$,  we obtain $M$ smeared fields $\hat{W}_n \deff \hat{\Phi}[f_n]$ at $M$  spatially separated points $\mathbf{x}_1, \mathbf{x}_2, \ldots , \mathbf{x}_M$ in the $xy$-plane.  We can then write
\begin{align}\label{a106}
\hat{W}_n = \sum_\mu \left( \hat{a}_\mu f_{n\mu}^* + \hat{a}_\mu^\dagger f_{n\mu} \right)/\sqrt{2},
\end{align}
where $f_{n\mu} = (u_\mu, f_n)$, and
\begin{align}\label{n107}
\sigma^2_n  = \left(f_n,f_n \right)/2, \qquad (n=1, \ldots, M).
\end{align}

\subsubsection{Particle-like operators}

We consider now the ``intensity'' operator   $\hat{\mathrm{I}}(\mathbf{x},z)$ defined by  $\hat{\mathrm{I}}(\mathbf{x},z) = \hat{\phi}^\dagger (\mathbf{x},z) \hat{\phi} (\mathbf{x},z)$. This quantity can be interpreted as a photon-number operator per unit transverse surface, because
\begin{align}\label{n110}
\int_{\mathbb{R}^2} \hat{\mathrm{I}}(\mathbf{x},z) \,  \mathrm{d} \mathbf{x} = \hat{N},
\end{align}
where $\hat{N}$ is defined by \eqref{n75}.

We then define the photon-counting operator $\hat{C}_n \deff \hat{\mathrm{I}}[\bm{1}_{n}], ~(n=1, \ldots, M)$ representing the action of detector $D_n$ when the latter is set to count the number of photons impinging on it, as
\begin{align}\label{a220}
\hat{C}_n \deff \hat{\mathrm{I}}[\bm{1}_{n}] = & \; \int_{\mathbb{R}^2}   \bm{1}_{n}(\mathbf{x}) \, \hat{\mathrm{I}}(\mathbf{x},z) \, \mathrm{d} \mathbf{x} \nonumber \\[6pt]
= & \; \bigl( \hat{\phi} , \bm{1}_{n} \hat{\phi} \bigr)\nonumber \\[6pt]
= & \; \sum_{\mu, \nu} \hat{a}_\mu^\dagger \hat{a}_\nu \bm{1}_{n\mu \nu},
\end{align}
where
\begin{align}\label{a222}
\bm{1}_{n \mu \nu} = (u_\mu, \bm{1}_{n} u _\nu).
\end{align}
By diagonalizing the linear operator whose matrix elements are $\bm{1}_{n \mu \nu}$, it is not difficult to find the discrete eigenvalues and eigenvectors of $\hat{C}_n$, but it is not necessary to do this. We need only to note that
by definition the \emph{discrete spectrum} of $\hat{C}_n$ gives the number of photons counted by detector $D_n$.   Note also that $\hat{C}_n$ is dimensionless.

\subsection{Commutation relations}

In the remainder we will need to use commutation relations for the wave and photon-counting operators $\hat{W}_n$ and $\hat{C}_n$. Such relations are calculated in Appendix \ref{commrel}, and the results are:
\begin{align}
\bigl[ \hat{W}_m, \, \hat{W}_n \bigr] &  = 0, \label{com10} \\[6pt]
\bigl[ \hat{C}_m, \, \hat{C}_n \bigr] &  = 0, \label{com20} \\[6pt]
\bigl[ \hat{W}_m, \, \hat{C}_n \bigr] &  = \frac{\delta_{nm}}{\sqrt{2}} \left\{ \bigl(f_n, \hat{\phi}  \bigr) - \bigl(\hat{\phi}, f_n \bigr) \right\}, \label{com30}
\end{align}
where $m,n = 1,2, \ldots, M$. Since all commutators above are zero for $m \neq n$, then all the wave an particle observables associated with different detectors are compatible and can be measured simultaneously.

\section{Probability distributions}\label{pd}

In random variable theory, the probability distribution or probability density function (p.d.f.) $p_\mathbf{Q}(\mathbf{q} )$ of a $M$-dimensional  random variable $\mathbf{Q} = (Q_1, Q_2, \ldots, Q_M)$, can be written as $p_\mathbf{Q}(\mathbf{q} ) = \langle \delta (\mathbf{Q} - \mathbf{q})\rangle$, where $\langle \cdots \rangle$ denotes average over all possible realization of $\mathbf{Q}$, and
\begin{align}\label{n130}
\delta (\mathbf{Q} - \mathbf{q}) = \prod_{n=1}^M \delta (Q_n - q_n),
\end{align}
 \cite{Ramshaw_1985}. Similarly, in quantum mechanics the spectral theorem \cite{aiello_arXiv.2110.12930} shows that given a Hermitian operator $\hat{Q}$ and a vector state $| \psi \rangle$ of norm $1$, we can  calculate the expectation value of any regular function $F(\hat{Q}) $ of $\hat{Q}$ with respect to $| \psi \rangle$, either as $\langle F(\hat{Q}) \rangle  = \langle \psi | F(\hat{Q}) | \psi \rangle$, or as
\begin{align}\label{n140}
\langle F(\hat{Q}) \rangle  = \int_{\mathbb{R}}  F(q) \, p_Q (q) \, \mathrm{d} q ,
\end{align}
where the p.d.f. $p_Q (q)$ of the random variable $Q$ associated with the operator $\hat{Q}$, is defined by
\begin{align}\label{n150}
p_Q (q)  = \langle \psi | \delta \bigl( \hat{Q} - q \bigr) | \psi \rangle,
\end{align}
(see, e.g., sec. 3-1-2 in \cite{ItzZub},  problem 4.3 in \cite{Coleman2019}, or  \cite{aiello_arXiv.2110.12930}). Using the Fourier representation of the Dirac delta function $\delta(z) = \int_\mathbb{R}  e^{i \alpha z} \, { \mathrm{d} \alpha }\, / (2 \pi)$, it is straightforward to show that
\begin{align}\label{a190}
p_Q (q) = \frac{ 1}{2 \pi} \int_\mathbb{R}  \langle \psi | e^{i \alpha \hat{Q}} | \psi \rangle e^{-i \alpha q}  \, \mathrm{d} \alpha  ,
\end{align}
where $\langle \psi | \exp({i \alpha \hat{Q}}) | \psi \rangle$ is the so-called quantum characteristic function \cite{MANKO1998328}.

The advantage of this formulation is that we can calculate $p_Q (q)$ without knowing the spectrum of $\hat{Q}$. Of course, if the latter were known, the calculation of $p_Q (q)$ would be trivial. To see this, suppose that $\hat{Q}$ is the position operator such that $\hat{Q} |q \rangle = q | q \rangle$. Then, from \eqref{n150} and a straightforward calculation it follows the well-known result
\begin{align}\label{n150bis}
p_Q (q)  = \left| \langle q | \psi \rangle \right|^2 = \left| \psi(q) \right|^2.
\end{align}

\section{Measuring the wave-like and particle-like aspects of light}\label{wp}

Consider three different experiments where a light beam prepared in the $N$-photon Fock state $| N[\phi] \rangle$ impinges upon a screen where two detectors are placed at two spatially separated points in the $xy$-plane, as shown in Fig. \ref{fig1}.
In the first experiment the detectors are set up to measures the  amplitudes $\mathcal{W}_1$ and $\mathcal{W}_2$ of the electric field of the light falling on them.
In the second experiment the detectors are set up to count the number of photons $\mathcal{C}_1$ and $\mathcal{C}_2$. Finally, in the third and last experiment detector $D_1$  measures the electric field  amplitude $\mathcal{W}_1$, and detector $D_2$ counts the number of photons $\mathcal{C}_2$.

The outcomes of these experiments can be described by the three pairs of  random variables $(W_1,W_2), ~ (C_1,C_2)$ and $(W_1,C_2)$, distributed according to
\begin{align}
(W_1,W_2) & \sim  p_{W_1 W_2}(N,w_1,w_2) , \nonumber \\[6pt]
(C_1,C_2) & \sim  p_{C_1 C_2}(N,c_1,c_2), \nonumber
\end{align}
and
\begin{align}
(W_1,C_2) & \sim    p_{W_1 C_2}(N,w,c), \nonumber
\end{align}
where either $N=0$ (vacuum state, we calculate it for comparison), or $N=1$ (single-photon state, the case of interest). Using the methods outlined in  Sec. \ref{pd}, these three p.d.f.s are calculated in Appendices \ref{pdfW}, \ref{pdfC} and \ref{pdfWC}, according to the formulas
\begin{multline}
p_{W_1 W_2}(N,w_1,w_2) \\[6pt]
 = \langle N[\phi] |  \delta \bigl( \hat{W}_1 - w_1 \bigr) \delta \bigl( \hat{W}_2 - w_2 \bigr) | N[\phi] \rangle, \label{a200}
\end{multline}
\begin{multline}
p_{C_1 C_2}(N,c_1,c_2) \\[6pt]
 = \langle N[\phi] |  \delta \bigl( \hat{C}_1 - c_1 \bigr) \delta \bigl( \hat{C}_2 - c_2 \bigr) | N[\phi] \rangle, \label{a210}
\end{multline}
and
\begin{multline}\label{a215}
p_{W_1 C_2}(N,w, c)  \\[6pt] = \langle N[\phi] | \delta \bigl( \hat{W}_1 - w \bigr)  \delta \bigl( \hat{C}_2 - c \bigr) | N[\phi] \rangle,
\end{multline}
where the $N$-photon Fock state $|N[\phi] \rangle$ is defined by  \eqref{n70}.
Note that there is not an operator ordering problem in the definitions \eqref{a200}-\eqref{a215} because all the operators involved do commute.

\subsection{Single detector}

For illustration purposes, here we  calculate the probability distributions of $W$ and $C$ alone, as if a single detector were present. In these two cases we have $W \sim p_{W}(N,w)$ and $C \sim p_{C}(N,c)$, where
\begin{align}
 p_{W}(N,w) &  = \langle N[\phi] | \delta \bigl( \hat{W} - w \bigr)  | N[\phi] \rangle, \label{one10} \\[6pt]
 p_{C}(N,c) &  = \langle N[\phi] | \delta \bigl( \hat{C} - c \bigr)  | N[\phi] \rangle, \label{one20}
\end{align}
with $N=0$ and $N=1$.

\subsubsection{Vacuum state}

In the simplest case  of input vacuum state, that is $N=0$, we have from \eqref{s365} and \eqref{s552}, $p_W(0,w) \deff p_0(w)$ and $p_C(0,c) \deff p_0(c)$, respectively, where
\begin{align}
 p_0(w) & =  \bigl( 2 \pi \sigma^2 \bigr)^{-1/2} \exp \left( -\frac{w^2}{2 \sigma^2} \right) , \label{a230} \\[6pt]
p_0 (c) & =  \delta(c), \label{a235}
\end{align}
and, as in \eqref{n105}, here $\sigma^2 = (f,f)/2$ fixes the variance of the smeared field $\hat{W}$ in the vacuum state.
As expected, $p_0 (w)$ and $p_0 (c)$ are the p.d.f.s of a continuous and a discrete random variable, respectively.

Equation \eqref{a230} shows a well-know result from the quantum theory of \emph{free fields}: the field amplitude in the ground (vacuum) state follows a Gaussian probability distribution that is centred on the value zero. The quantum field fluctuations are fixed by the smearing function via $\sigma^2$.
For example, if $f(\mathbf{x})$ is given by \eqref{a172}, then $\sigma^2 = 1/(4 \pi a^2)$. This implies that when the linear dimension $a$ of the region in which the field amplitude is measured shrinks to zero, the quantum fluctuations become huge, eventually becoming infinite for $a \to 0$ \cite{Coleman2019}.

Equation \eqref{a235} gives the trivial p.d.f. of a discrete-type random variable with a probability mass function that takes a single value: $\operatorname{Prob}(C = 0) = 1$. Physically this means that in the vacuum state of the electromagnetic field, the probability to count one or more photons is equal to zero, as it should be.

\subsubsection{Single-photon state}

Next, we write down the probability distributions \eqref{one10} and \eqref{one20} with respect to the single-photon state ($N=1$). From \eqref{q640} and \eqref{s735}, we have
\begin{align}
p_W (1,w)  & =  \bigl( 1 - |s|^2 \bigr)p_0 (w) + |s|^2 p_1 (w) ,\label{a240} \\[6pt]
p_C (1,c)  & =   ( 1-P ) \, \delta(c) + P \, \delta(c-1) , \label{a250}
\end{align}
where $p_0 (w)$ and $p_1 (w) = p_0 (w) w^2/\sigma^2$ are given by \eqref{a230} and  \eqref{q650}, respectively. Moreover, we have set
\begin{align}\label{n200}
s = \frac{(\phi,f)}{(f,f)^{1/2}} , \quad \text{and} \quad P = (\phi, \bm{1} \phi) \geq 0,
\end{align}
with $\bm{1}(\mathbf{x})$  denoting the indicator function of the domain $\mathcal{D}$ representing the active area of the single detector. As usual, $f$ denotes the smearing function.
The physical meaning of the two key parameters $s$ and $P$ is the following. By definition, $\tilde{f} \deff f/(f,f)^{1/2}$ is normalized to $(\tilde{f},\tilde{f}) = 1$, as is $(\phi,\phi) = 1$, due to \eqref{n55}.
Therefore, 
from the field-quadrature interpretation \eqref{a170bis} of the wave operator $\hat{W}$, it follows that $s = (\phi, \tilde{f})$ quantifies the mode-matching between the single-photon field $\phi$ and the spatial mode  $\tilde{f}$  of the supposed local oscillator that would perform the homodyne measurement of the quadrature  $\hat{W}$. The better the matching, the greater the value of $|s|^2$.
The parameter $P$, here rewritten as
\begin{align}\label{p10}
P = \int_{\mathcal{D}} \left| \phi(\mathbf{x},z) \right|^2 \, \mathrm{d} \mathbf{x},
\end{align}
gives the fraction of the intensity $\left| \phi(\mathbf{x},z) \right|^2$ of the incident beam that falls upon the detector surface.

From \eqref{a240} it is easy to calculate the average value $\operatorname{E}[W]=0$,
and the variance $\sigma_W^2$ of $W$,
\begin{align}\label{n203bis}
\sigma_W^2 = \operatorname{E}[W^2] - (\operatorname{E}[W])^2 = \sigma^2 (1 + 2 |s|^2).
\end{align}
Equation \eqref{n203bis} shows that the quantum fluctuations of the field  in the single-photon state are always bigger that the fluctuations in vacuum.
Similarly, from \eqref{a250} it follows that
\begin{align}\label{p12}
\operatorname{E}[C] = P = \operatorname{E}[C^2],
\end{align}
so that the variance $\sigma_C^2$ of $C$ is
\begin{align}\label{p14}
\sigma_C^2 = \operatorname{E}[C^2] - (\operatorname{E}[C])^2 = P(1-P).
\end{align}

\begin{figure}[ht!]
  \centering
  \includegraphics[scale=3,clip=false,width=1\columnwidth,trim = 0 0 0 0]{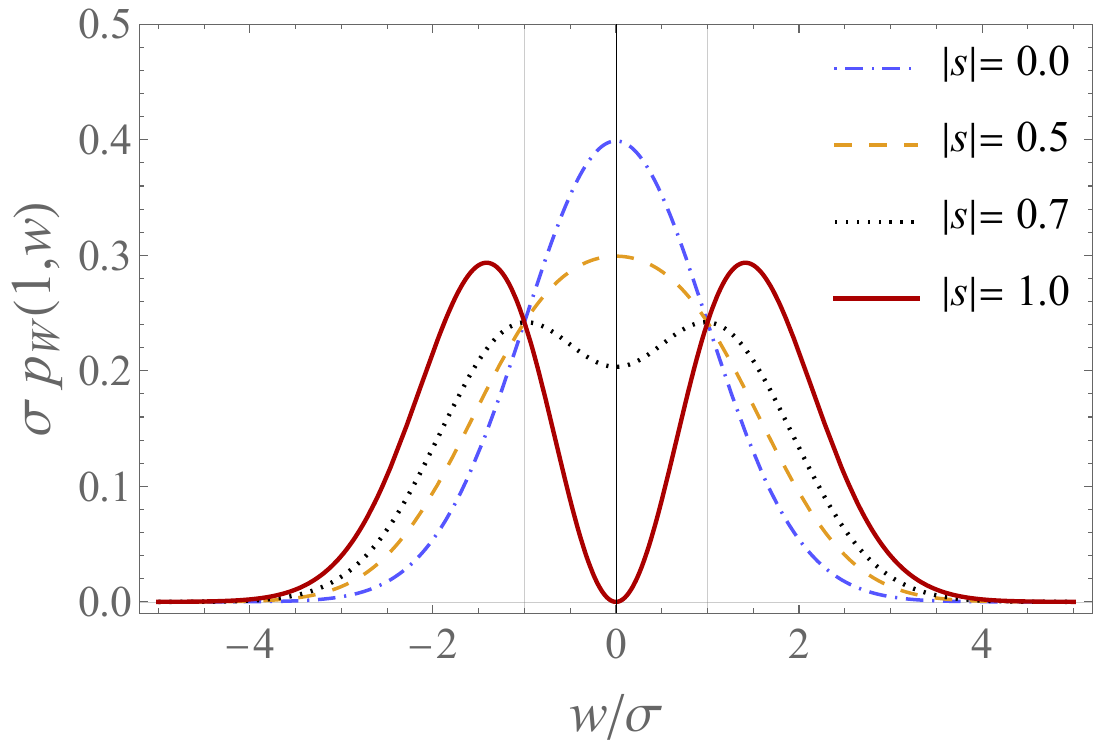}
  \caption{Plots of $p_W (1,w)$ given by \eqref{a240} for different values of $|s| \in [0,1]$. The blue dot-dashed curve for $|s|=0$ is equal to the p.d.f. for the vacuum state $p_W (0,w)$. When the superposition between the cross section of the beam and the detector surface increases, the central value  $p_W (1,0)$ decreases to zero. Note that from \eqref{a240} it directly follows that all the curves plotted here intersect at $w=\sigma$.}\label{fig2}
\end{figure}

Equation \eqref{a240} shows that $p_W(1,w)$ is a so-called \emph{mixture distribution}  (see, e.g., Sec. \textbf{3.5} of Ref. \cite{MurphyBook2022}), that is a convex combination of the elementary distributions $p_0 (w)$ and $p_1 (w)$, with weights $1 - |s|^2$ and $|s|^2$, respectively. From a physical point of view, this means that when we measure some amplitude $w$ of the field, there is a probability $1 - |s|^2$ that this amplitude was sampled from the ``vacuum distribution'' $p_0 (w)$, and a probability $|s|^2$ that it was sampled from the ``single-photon distribution'' $p_1 (w)$, instead. This ambiguity can be reduced in one direction or the other by varying $|s|$. When $|s| = 0$ only the vacuum state contribute to $p_W (1,w)$. This is clear. However, if
$|s| =1$, the contribution of the vacuum to $w$ will be zero, and
\begin{align}\label{n202}
p_W (1,w) =   p_1 (w)  =   p_0 (w)  \frac{w^2}{\sigma^2} .
\end{align}
This kind of distribution is known as the  Maxwell distribution of speeds in statistical physics when $w \geq 0$ (see, e.g., Sec. 7.10 in \cite{reif2009}). Note that at $w=0$
we have $p_W (1,0) = p_0 (0)(1 - |s|^2)$, with $1 - |s|^2 \leq 1$ from \eqref{n200} and the Cauchy–Schwarz inequality. This $s$-dependent dip at $w=0$, is the signature of the single-photon state with respect to the vacuum state, as illustrated by Fig. \ref{fig2}.

The p.d.f. \eqref{a250} of the discrete random variable $C$ gives $\operatorname{Prob}(C = 0) = 1 - P$, and $\operatorname{Prob}(C = 1) = P$, where $P$ is given by \eqref{n200}.

\subsection{Two detectors}

When there are two detectors located in two different places in the $xy$-plane, as shown in Fig. \ref{fig1}, we can choose between three different possibilities of detection: \emph{a}) wave-wave detection; \emph{b}) particle-particle detection; and \emph{c}) wave-particle detection. In the remainder of this section  we will analyze in detail these three cases.

\subsubsection{a) Wave-wave detection}

For the vacuum state and $M=2$, \eqref{s366}  gives
\begin{align}\label{n220}
p_{W_1 W_2}  (0,w_1,w_2) &  =   p_0 (w_1) \, p_0 (w_2) \nonumber \\[6pt]
& \deff p_0(w_1,w_2) ,
\end{align}
where $ p_0 (w)$ is  defined by \eqref{a230}. Thus, the joint p.d.f. is the product of the marginal probability distributions $p_0 (w_1) $ and $p_0 (w_2) $. Therefore, the two random variables $W_1$ and $W_2$ defined by \emph{both} the operators $\hat{W}_1, ~\hat{W}_2$ and the quantum vacuum state $| 0 \rangle$, are independent.

\begin{figure}[ht!]
  \centering
  \includegraphics[scale=3,clip=false,width=1\columnwidth,trim = 0 0 0 0]{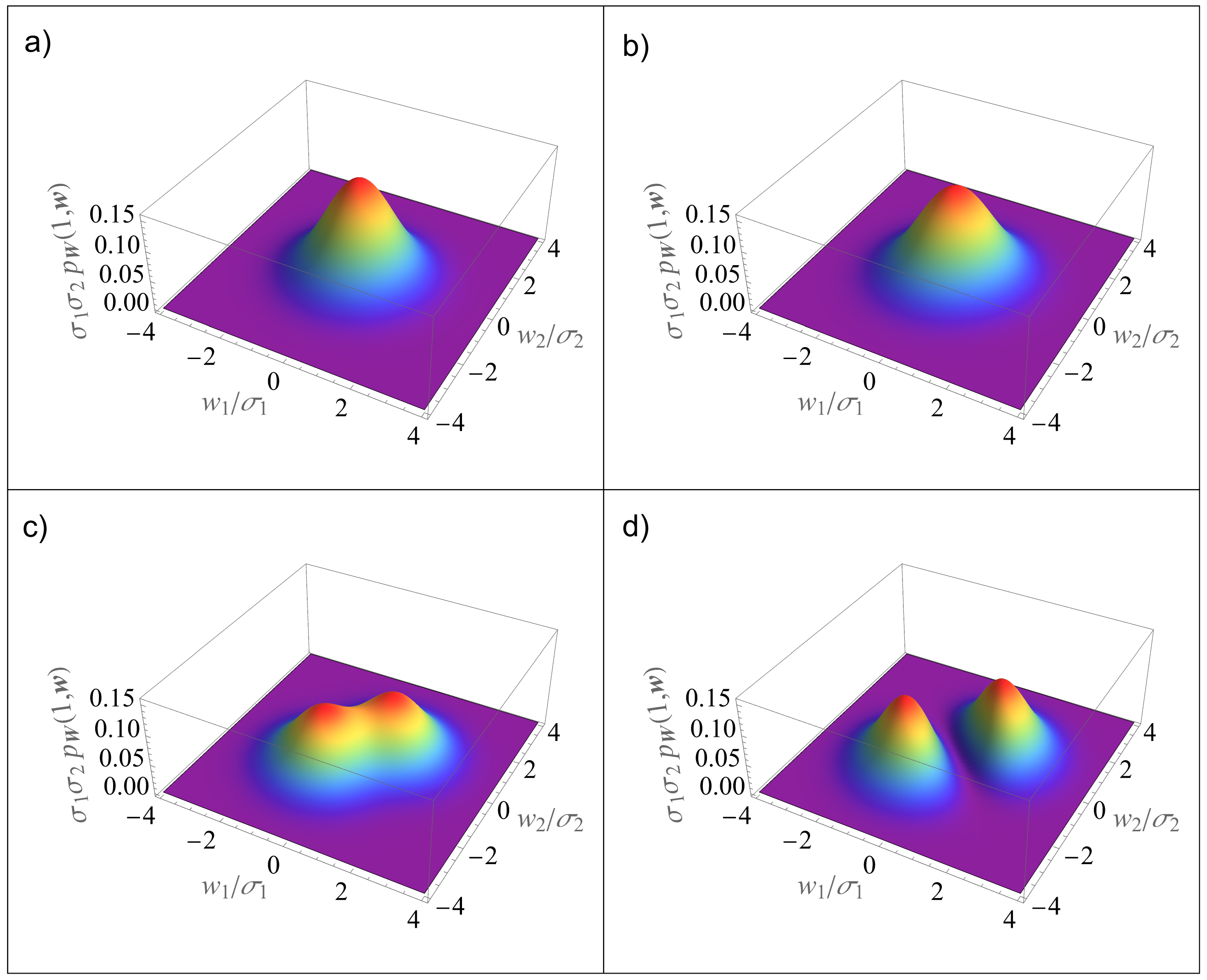}
  \caption{Plots of $p_{W_1 W_2} (1,w_1,w_2) \deff p_{\mathbf{W}} (1,\mathbf{w})$ given by \eqref{a270} for $s_1 = s_2 = s \in \mathbb{R}$, and: \textsf{a)} $s=0$, (equal to the p.d.f. for the vacuum state);  \textsf{b)} $s=0.2$; \textsf{c)} $s=0.5$; \textsf{d)} $s=0.7$. The maximum permitted value for $s$ is $s = 1/\sqrt{2} \approx 0.71$. Note the ``dip'' at $w_1/\sigma_1 + w_2/\sigma_2 =0$, analogous to the one occurring for the single-detector case.}\label{fig3}
\end{figure}

For the single-photon state $|1[\phi] \rangle$, equations \eqref{q610}-\eqref{q620} give
\begin{align}\label{a270}
p_{W_1 W_2}  (1,w_1,w_2) & = \bigl( 1 - |\mathbf{s}|^2 \bigr) \, p_0(w_1,w_2) \nonumber \\[6pt]
& \phantom{=}  +   |\mathbf{s}|^2 \, p_1(w_1,w_2),
\end{align}
which is, as in the single-detector case,  a mixture distribution.
In this equation $\mathbf{s} = (s_1,s_2)$, with $|\mathbf{s}|^2 = |s_1|^2 + |s_2|^2$, and
\begin{align}\label{a270bis}
p_1(w_1,w_2) = p_0(w_1,w_2)\left| \frac{w_1}{\sigma_1}\, \frac{s_1}{|\mathbf{s}|} + \frac{w_2}{\sigma_2} \, \frac{s_2}{|\mathbf{s}|} \right|^2,
\end{align}
where
\begin{align}\label{n290}
\sigma_n^2 = \frac{(f_n,f_n)}{2}, \qquad (n = 1,2),
\end{align}
and
\begin{align}\label{n292}
s_n = \frac{(\phi,f_n)}{(f_n,f_n)^{1/2}} = \frac{(\phi,f_n)}{\sqrt{2} \,\sigma_n}, \qquad (n = 1,2).
\end{align}
By definition $|s_n| \leq 1$. However, the existence of \eqref{a270} imposes the further joint condition $|s_1| + |s_2| \leq 1$.
A pictorial representation of the distribution $p_{W_1 W_2}  (1,w_1,w_2)$ is shown in  Fig. \ref{fig3}.

The set of parameters that characterize the bivariate distribution \eqref{a270} can be straightforwardly calculated. The average values are equal to zero, i.e.,  $\operatorname{E}[W_1]=0 = \operatorname{E}[W_2]$. The  variances are:
\begin{align}\label{ww10}
\sigma_{W_n}^2 \deff \operatorname{E}[W_n^2]= \sigma_n^2 (1 + 2 |s_n|^2),
\end{align}
with $n=1,2$, and the covariance is
\begin{align}\label{ww20}
\operatorname{E}[W_1 W_2] = \sigma_1 \sigma_2 \left(s_1 s_2^* + s_1^* s_2 \right).
\end{align}

If we choose the smearing functions $f_1$ and $f_2$ and the location of the two detectors such that $\sigma_1=\sigma_2 \deff \sigma$ and $s_1=s_2 \deff s$, then using \eqref{ww10}-\eqref{ww20} we find the following correlation coefficient:
\begin{align}\label{ww30}
\frac{\operatorname{E}[W_1 W_2] - \operatorname{E}[W_1] \operatorname{E}[W_2]}{ \sigma_{W_1}  \sigma_{W_2}} & = \frac{2 \,  |s |^2}{1 + 2 \,  |s |^2}  \nonumber \\[6pt]
& \leq \frac{1}{2}.
\end{align}
The latter inequality follows from the condition $0 \leq 1 - |s_1|^2 - |s_2|^2$, which becomes $|s|^2 \leq 1/2$ in the present case.
A positive correlation coefficient between $W_1$ and $W_2$ means that when $W_1$ increases then $W_2$ also increases, and when $W_1$ decreases then $W_2$ also decreases. The minimum value $0$ of the correlation coefficient \eqref{ww30} is attained when $s=0$. This may occur in two different ways: either \emph{a}) both detectors have finite active surface but are located outside the section of the beam on the detection screen, or \emph{b}) the detectors are placed within the section of the beam, but they are point-like detectors with zero-size active surface.
Case  \emph{a}) is trivial  and implies $p_{W_1 W_2}  (1,w_1,w_2) = p_0(w_1,w_2) $.
Case \emph{b}) is more interesting because it shows that the amplitudes of the field measured by any pair of point-like detectors are always uncorrelated. This is due to the fact that a quantum field wildly fluctuates  when it is strongly localized. To see this, let us take  $f(\mathbf{x})$ as in \eqref{a172}, so that $(f,f) = 1/(2 \pi a^2) = 2 \sigma^2$. Then, to achieve $s_1=s_2=s$ we must assume that  the two point-like detectors are located at  $\mathbf{x}_1$ and $\mathbf{x}_2$ chosen in such a way that $|\phi(\mathbf{x}_1,z)| = |\phi(\mathbf{x}_2,z)|$.  In this case from \eqref{n290} and $a \approx 0$, it follows that
\begin{align}\label{ww35}
|s|^2  = \frac{|(\phi,f)|^2}{2 \, \sigma^2}  \approx \sqrt{2 \pi} \, a \, |\phi(\mathbf{x}_1,z)|^2.
\end{align}
Since $|\phi(\mathbf{x}_1,z)|^2$ is always a finite quantity for any physically realisable light beam, then $|s|^2 \to 0$ when the size $a$ of the detectors goes to zero. But when $a \to 0$, the quantum fluctuations blow up because $\sigma^2 = 1/(4 \pi a^2) \to \infty $.

It is interesting to note that when either $s_1=0$ or $s_2 = 0$,  the two random variables $W_1$ and $W_2$ become independent. However, when both $s_1 \neq 0$ and $s_2 \neq 0$, then $W_1$ and $W_2$ are not independent although the two corresponding operators $\hat{W}_1$ and $\hat{W}_2$ do commute. This is a consequence of the  non-localizability of the single-photon field $\phi(\mathbf{x},z)$, whose section  in the $xy$-plane extends over both regions $\mathcal{D}_1$ and $\mathcal{D}_2$ covered by the active surfaces of the two detectors $D_1$ and $D_2$. In fact, the joint p.d.f. $p_{W_1 W_2}  (1,w_1,w_2)$ is  determined by \emph{both} the operators $\hat{W}_1, ~\hat{W}_2$ \emph{and} the quantum state $|1[\phi]\rangle$. Therefore, it is the spatial transverse extension of the light field $\phi(\mathbf{x},z)$ that establishes a correlation between the two random variables $W_1$ and $W_2$.
Very similar conclusions were reached in Ref. \cite{PhysRevLett.92.193601} where the authors investigated, in their own words, ``the delocalized state formed by a photon'' using homodyne tomography.

\subsubsection{ b) Particle-particle detection}\label{pp}

For the vacuum state and $M=2$, \eqref{s550}  gives
\begin{align}\label{n230}
p_{C_1 C_2}  (0,c_1,c_2)  =   \delta(c_1) \, \delta(c_2).
\end{align}
 This equation simply shows that in the vacuum state we always count zero photons.

More interesting is the single-photon state case, for which   \eqref{s720}  gives
\begin{widetext}
\begin{align}\label{a280}
p_{C_1 C_2} (1,c_1,c_2)    =  (1 - P_1 - P_2) \, \delta(c_1) \, \delta(c_2)  + P_1 \, \delta(c_1-1) \, \delta(c_2) +  P_2 \, \delta(c_1) \, \delta(c_2-1) ,
\end{align}
\end{widetext}
where
\begin{align}\label{n290bis}
P_n = (\phi, \bm{1}_n \phi) \leq 1, \qquad (n = 1,2),
\end{align}
is the fraction of the intensity of the beam impinging upon the $n \text{th}$ detector. Note that the first term in \eqref{a280} enforces the constraint $P_1 + P_2 \leq 1$.
Using \eqref{a280} it is not difficult to calculate
\begin{align}\label{cc10}
\operatorname{E}[(C_n)^k]= P_n, \qquad (k \in \mathbb{N}),
\end{align}
with $n=1,2$, and
\begin{align}\label{cc20}
\operatorname{E}[C_1 C_2] = 0.
\end{align}
The latter two equations imply for the correlation coefficient,
\begin{align}\label{cc30}
\frac{ \operatorname{E}[C_1 C_2] - \operatorname{E}[C_1] \operatorname{E}[C_2]}{\sigma_{C_1} \sigma_{C_2}} &  = - \sqrt{\frac{P_1 P_2}{(1-P_1)(1-P_2)}}\nonumber \\[6pt]
& \geq -1 ,
\end{align}
where \eqref{p14} has been used.
A negative correlation means that there is an inverse relationship between the random variables $C_1$ and $C_2$. In physical terms this means that  when the number of photons counted by $D_1$ increases, the one counted by $D_2$ must decrease, and vice versa. This is a consequence of both the fixed  the number of photons in Fock states and  the non-localizability of the electromagnetic field, which we have previously discussed. Differently from \eqref{ww30}, here the  correlation coefficient can achieve the minimum value $-1$, which means perfect anticorrelation between $C_1$ and $C_2$. This occurs when $P_1=P_2 = 1/2$, which means that we are using a split detector to count photons in each half of the beam. In this case, the first term in \eqref{a280} (the vacuum contribution), goes to zero.

\subsubsection{c) Wave-particle detection}\label{wpd}

This is the last and more interesting case. By hypothesis, detector $D_1$  measures the electric-field amplitude $\mathcal{W}_1$ of a portion of the impinging light beam, and detector $D_2$ counts the number of photons $\mathcal{C}_2$ in a different portion of the same beam.
For the vacuum state \eqref{d20}  gives
\begin{align}\label{wp230}
p _{W_1C_2}  (0,w,c) =  p_0(w) \, \delta(c),
\end{align}
as expected. This result is very simple and there is not much to say about it.

\begin{figure}[hb!]
  \centering
  \includegraphics[scale=3,clip=false,width=1\columnwidth,trim = 0 0 0 0]{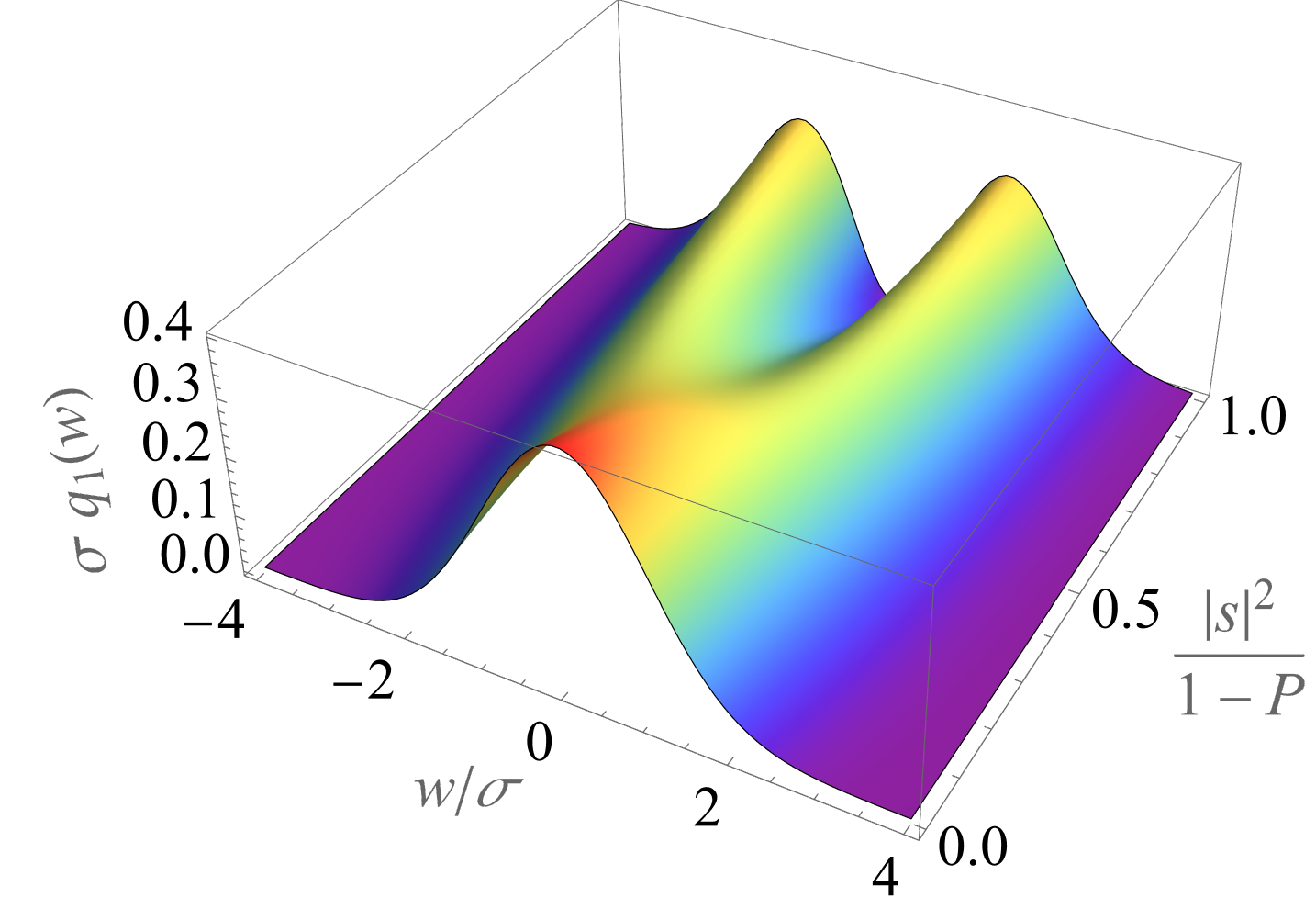}
  \caption{Plot of the probability distribution $q_1(w)$ given by \eqref{wp10bis}. By definition
  $0 \leq |s|^2/(1-P) \leq 1$, because $|s|^2 \to 0$ when $P \to 1$. When $|s|^2/(1-P) =0$, we recover the vacuum-field distribution, that is $q_1(w) = p_0(w)$. In contrast, when $|s|^2/(1-P) = 1$, we have $q_1(w) = p_1(w)$, where $p_1(w)$ is defined by \eqref{n202}. In this limit $q_1(w)$ receives no contribution from the vacuum field, and the  dip at $w=0$ is again the hallmark of the single-photon field.}\label{fig4}
\end{figure}

Conversely, for the single-photon state from \eqref{d70} we have,
\begin{align}\label{wp10}
p_{W_1 C_2}  (1,w,c)  & = \bigl( 1 - P \bigr)\, \delta(c) \, q_1(w) \nonumber \\[6pt]
& \phantom{=} + P \, \delta(c-1) \, p_0(w) ,
\end{align}
where $p_0 (w)$ is given by \eqref{a230}, $s$ and $P$ are defined by \eqref{n200}, and the mixed distribution $q_1(w)$ is defined by
\begin{align}\label{wp10bis}
q_1(w)  & \deff  \left(1-\frac{|s|^2}{1-P} \right)   p_0(w) \nonumber \\[6pt]
& \phantom{\deff}  + \frac{|s|^2}{1-P} \, p_1(w),
\end{align}
 where $p_1(w)$ is defined by \eqref{n202}. Note that $q_1(w)$ coincides with $p_W(1,w)$ given by \eqref{a240}, if in the latter we replace $|s|^2$ with $|s|^2/(1-P)$.
 The first term in  \eqref{wp10bis} is due to the  vacuum-field fluctuations, while the second term accounts for the $\phi$-dependent contribution from the single-photon field.
This distribution is illustrated in Fig. \ref{fig4}.
\\
The meaning of the two terms in \eqref{wp10} should be clear. The first term tells us that there is a probability $1-P$ that  the pair of observables $(\mathcal{W}_1, \mathcal{C}_2)$ takes the values $(w,0)$, with the wave amplitude $w$ being sampled either from the vacuum-field distribution $p_0(w)$ with probability $1 -|s|^2/(1-P)$, or from the single-photon distribution $p_1(w)$ with probability ${|s|^2}/(1-P)$.  The second term shows that whenever we measure the pair of values $(\mathcal{W}_1, \mathcal{C}_2) = (w,1)$, then the value of $w$ has  been sampled from the vacuum distribution with certainty. This demonstrates the mutually exclusively dual nature, discrete and continuous, of the single-photon field.

The first two moments that characterize the p.d.f. \eqref{wp10}, are
\begin{align}\label{wp20}
\operatorname{E}[W_1] = 0, \qquad   \operatorname{E}[C_2] = P,
\end{align}
and
\begin{align}
\operatorname{E}[W_1^2] & = \sigma^2 \left( 1+ 2 |s|^2\right), \label{wp30} \\[6pt]
\operatorname{E}[C_2^2] & = P, \label{wp40} \\[6pt]
\operatorname{E}[W_1 C_2] & = 0. \label{wp50}
\end{align}
This implies
\begin{align}
\sigma_{W_1} & = \sqrt{\operatorname{E}[W_1^2] - (\operatorname{E}[W_1])^2} = \sigma^2 ( 1+ 2 |s|^2 ), \label{wp52A} \\[6pt]
\sigma_{C_2} & = \sqrt{\operatorname{E}[C_2^2] - (\operatorname{E}[C_2])^2} = \sqrt{P(1-P)}, \label{wp52B}
\end{align}
with $0 \leq P \leq 1$.
Therefore, from \eqref{wp20}-\eqref{wp52B} it follows that the \emph{linear} correlation coefficient of $W_1$ and $C_2$ is zero:
\begin{align}\label{wp55}
 \frac{\operatorname{E}[W_1 C_2] - \operatorname{E}[W_1] \operatorname{E}[C_2]}{\sigma_{W_1} \, \sigma_{C_2}}  = 0 .
\end{align}

Interestingly, the positivity of the  first term in the distribution \eqref{wp10bis} implies the condition $1 - |s|^2 - P \geq 0$, which results in the inequality
\begin{align}\label{wp15}
|s|^2 + P \leq 1.
\end{align}
This simple expression somehow quantifies wave-particle duality  in that it establishes a connection between the probability $|s|^2$ that the photon reaches the wave detector $D_1$, thus revealing its wave-like nature, and the probability $P$ that it hits the particle detector $D_2$, then manifesting its particle-like character.

\section{Discussion of the results}\label{discuss}

\subsection{Comparison of the linear correlations}\label{compalin}

To begin with, let us compare the linear correlation coefficients of $(W_1,W_2)$, $(C_1,C_2)$, and $(W_1,C_2)$, given by \eqref{ww30}, \eqref{cc30}, and \eqref{wp55}, respectively, that we rewrite here as
\begin{align}\label{ww30bis}
\rho_{W_1 W_2} \deff \frac{\operatorname{E}[W_1 W_2] - \operatorname{E}[W_1] \operatorname{E}[W_2]}{ \sigma_{W_1}  \sigma_{W_2}}  \leq \frac{1}{2},
\end{align}
\begin{align}\label{cc30bis}
\rho_{C_1 C_2} \deff \frac{ \operatorname{E}[C_1 C_2] - \operatorname{E}[C_1] \operatorname{E}[C_2]}{\sigma_{C_1} \sigma_{C_2}}  \geq -1 ,
\end{align}
and
\begin{align}\label{wp55bis}
\rho_{W_1 C_2} \deff  \frac{\operatorname{E}[W_1 C_2] - \operatorname{E}[W_1] \operatorname{E}[C_2]}{\sigma_{W_1}  \sigma_{C_2}}  = 0 .
\end{align}

In the wave-wave detection case the two ``wavelike'' random variables $W_1$ and $W_2$ are linearly (positively) correlated, but their the correlation coefficient $\rho_{W_1 W_2}$ can not reach the maximum value $1$, even in the best case scenario with $\sigma_1=\sigma_2 \deff \sigma$ and $|s_1|^2=|s_2|^2= 1/2$, when the vacuum contribution to $p_{W_1,W_2}(1,w_1,w_2)$ goes to zero. The constraint $\rho_{W_1 W_2} \leq \frac{1}{2}$ follows from the rightmost ``interference term'' in \eqref{a270bis}, yielding to \eqref{ww20}. This term  causes a dispersion of the values of $w_1$ and $w_2$ around the two peaks of the probability distribution $p_{W_1 W_2}(1, w_1,w_2)$, as shown in Fig. \ref{fig3} \textsf{d)}, and it is due to the non-local character of the single-photon field that covers both detectors, as previously discussed.

Vice versa, in the particle-particle case  the (singular) probability distribution ${p_{C_1 C_2}(1, c_1,c_2)}$ is strictly localized in the $c_1c_2$-plane around the three points $(c_1,c_2) = (0,0)$, $(c_1,c_2) = (0,1)$ and $(c_1,c_2) = (1,0)$. In idealized experimental conditions where the two  detectors have $100\%$ efficiency and intercept completely the light beam (that is, $P_1 = P_2 = 1/2$), the first point $(c_1,c_2) = (0,0)$ cannot occur thus yielding perfect anticorrelation between $C_1$ and $C_2$, that is  $\rho_{C_1 C_2} = -1$.
In other words, it is the localized nature of the single-photon detection process that yields maximal anticorrelation.

Finally, in the wave-particle detection case the continuous-discrete mixture probability distribution $p_{W_1 C_2} (1,w,c)$ generates two sets of sample points that lay in the $wc$-plane on the lines $c=0$ and $c=1$, respectively. The points on the line $c=0$ are sampled from the (single-photon) distribution $q_1(w)$, while the points on the line $c=1$ are generated by the  (vacuum) distribution $p_0(w)$, as implied by \eqref{wp10}. From a physical point of view, the two distributions  $q_1(w)$ and  $p_0(w)$ differs because of the local nature of the single-photon detection process (a photon cannot be split in two): when the photon is measured in the part of the field intercepted by the wave-detector $D_1$, the field values follow the distribution $q_1(w)$. Vice versa, when the photon falls on the particle-detector $D_2$, the field amplitudes measured by $D_1$ are due solely to  vacuum field fluctuations. From a mathematical point of view,  $q_1(w) \neq p_0(w)$ implies that the probability density function $p_{W_1 C_2}(1,w,c)$ is not separable with respect to the variables $w$ and $c$. The physical counterpart of this statement is that the random variables $W_1$ and $C_2$ are \emph{not} independent, that is
\begin{align}\label{nofact}
p_{W_1 C_2}(1,w,c) \neq p_{W_1}(1,w)p_{C_2}(1,c),
\end{align}
where the marginal distributions $p_{W_1}(1,w)$ and $p_{C_2}(1,c)$ are given by \eqref{a240} and \eqref{a250}, respectively.
This implies, surprisingly enough, that there exist \emph{nonlinear} (quadratic, cubic, etc.,) correlations between the wave-like (continuous) random variable $W_1$ and the particle-like  (discrete) one $C_2$. In fact, it is not difficult to show that
the first nonzero correlation coefficient is given by the \emph{quadratic} (with respect to $W_1$), central moment of the wave-particle distribution $p _{W_1 C_2}  (1,w,c)$,  that is
\begin{align}\label{wp57}
\operatorname{E}\left[\bigl( W_1^2 - \operatorname{E}[W_1^2] \bigr)(C_2 - \operatorname{E}[C_2]) \right] = - 2 \sigma^2 |s|^2 P.
\end{align}
Thus,  \eqref{wp55bis} and \eqref{wp57} show that there is no \emph{linear} dependence between  the wave and particle random variables $W_1$ and $C_2$, but there is at least a quadratic one \cite{MurphyBook2022}. Now we will describe a practical way to quantify such nonlinear dependence.

\subsection{Quantifying the nonlinear dependence: The mutual information}\label{mutual}

It is well known in probability theory that  when random variables are correlated in a non-linear manner one must use more suitable measures of dependence such as the mutual information \cite{CoverThomas}.
Mutual information of $W_1$ and $C_2$ tells how different the joint distribution $p _{W_1  C_2}  (1,w,c)$  is, from the product of the marginal distributions $p_{W_1}(1,w)$ and $p_{C_2}(1,c)$.
In practice, mutual information quantifies the reduction in the average uncertainty about one random variable given the knowledge of another.
Thus, large values of mutual information indicate high reduction in uncertainty; small values of mutual information denote low reduction; and zero mutual information  means that the two random variables are independent.
We stress that here the term ``uncertainty'' is referring to the values taken by the random variables $W_1$ and $C_2$, and it should not be confused with the well-known Heisenberg uncertainty, which refers to non-compatible, conjugate observables, which, differently from $\mathcal{W}_1$ and $\mathcal{C}_2$, cannot be measured simultaneously. Clearly, for conjugate  observables a joint probability distribution cannot be calculated and, consequently, mutual information cannot be defined.\footnote{However,  using Wigner's functions and \emph{linear entropy}, a different form of mutual information can be defined also for non-compatible observables \cite{PhysRevE.62.4665,SANTOS2021125937}.}

\begin{figure}[ht!]
  \centering
  \includegraphics[scale=3,clip=false,width=1\columnwidth,trim = 0 0 0 0]{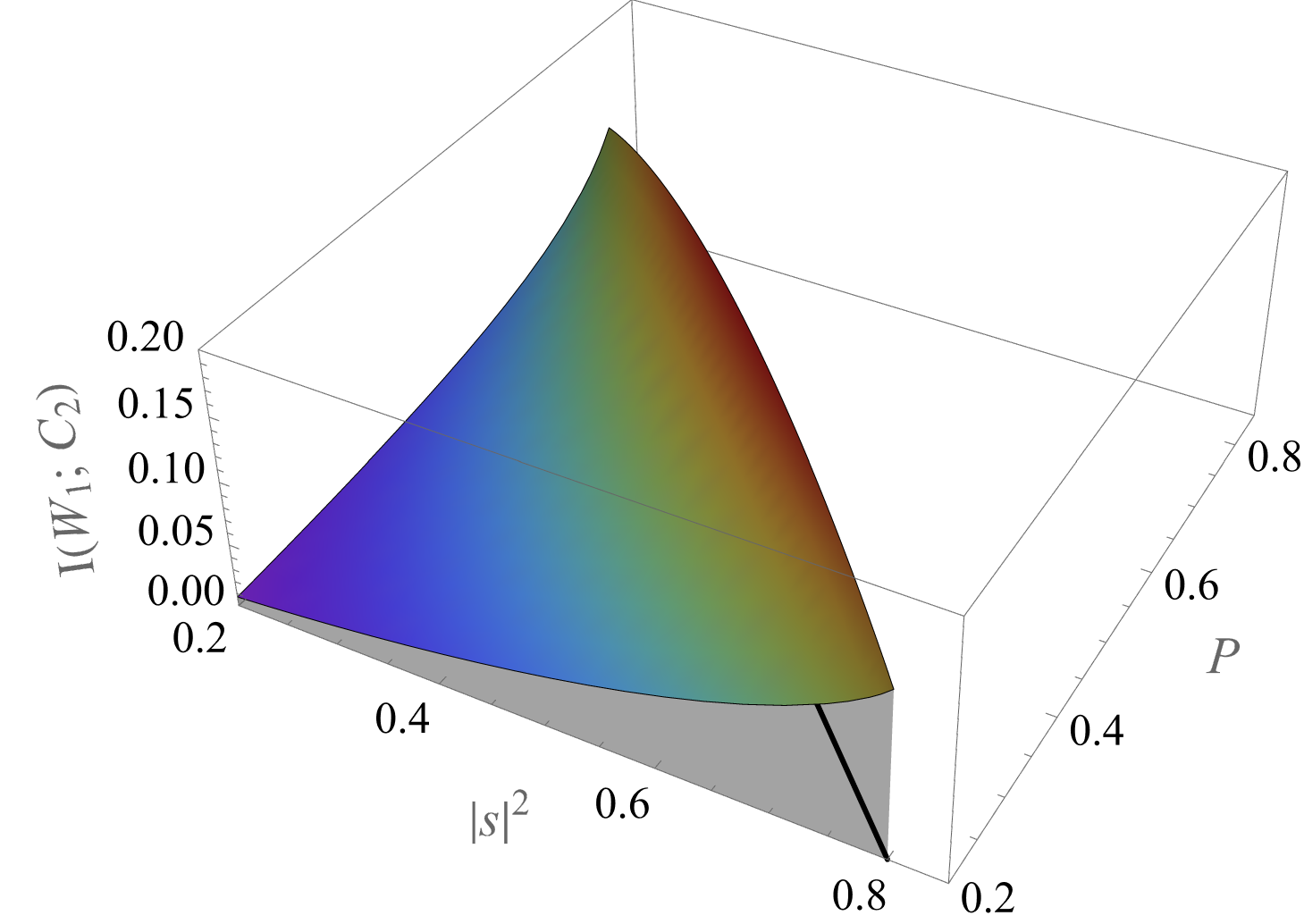}
  \caption{Plot of $\operatorname{I}(W_1;C_2)$ given by \eqref{wp70}. Note that the domain of the function is the region of the $|s|^2P$-plane defined by $1 -|s|^2 -P \geq 0$ (gray area in the plot).   For point-like detectors  we have $P , |s|^2 \ll 1$, so that  $\operatorname{I}(W_1;C_2) \approx P |s|^4/(1-P)$.}\label{fig5}
\end{figure}

For a mixture of discrete and continuous variables the mutual information can be written as \cite{nair2007entropy},
\begin{align}\label{wp70}
\operatorname{I}(W_1;C_2)   = \operatorname{h}(W_1) + \operatorname{H}(C_2) - \mathcal{H}(W_1,C_2 ),
\end{align}
where we have defined the continuous (differential), discrete and mixed entropies, $\operatorname{h}(W_1)$, $\operatorname{H}(C_2)$, and  $\mathcal{H}(W_1,C_2)$, respectively, as
\begin{align}
\operatorname{h}(W_1) & = -\int_\mathbb{R} p_{W_1}(1,w) \ln \bigl[ p_{W_1}(1,w) \bigr] \, \text{d}w, \label{wp80} \\[6pt]
\operatorname{H}(C_2) &  = -P \ln P - (1-P) \ln (1-P), \label{wp90}
\end{align}
and
\begin{align}\label{wp100}
\mathcal{H}(W_1 , C_2)   = - \sum_{i=0}^1 \int_\mathbb{R} g_i(w) \ln \bigl[g_i(w) \bigr] \, \text{d} w.
\end{align}
The two functions $g_0(w)$ and $g_1(w)$ are defined by
\begin{align}\label{wp110}
g_0(w) \deff (1-P)q_1(w), \quad g_1(w) \deff P p_0(w),
\end{align}
where $q_1(w)$ is given by \eqref{wp10bis}.
The quantities in \eqref{wp80}-\eqref{wp100} can be calculated explicitly, for example by using Mathematica \cite{Mathematica}. We do not write down the formulae here as they are very complicated and not particularly enlightening. However, we plot $\operatorname{I}(W_1;C_2)$ in Fig. \ref{fig5}.
The existence of a nonzero mutual information witnesses the presence of a nonlinear relationship between the wave and the particle observables $\mathcal{W}_1$ and $\mathcal{C}_2$, respectively. This is the main result of this paper.

The maximum value of the mutual information is achieved for $P = 1 - |s|^2 \approx 0.47$, and it is given by  $\max [\operatorname{I}(W_1;C_2)] \approx 0.18$. To understand what this number means, we compare  the maximum of $\operatorname{I}(W_1;C_2)$ with the maximum of $\operatorname{I}(C_1;C_2)$, the latter being given by
\begin{align}\label{wp120}
\operatorname{I}(C_1;C_2) & = -(1-P_1)\ln(1-P_1) \nonumber \\[6pt]
& \phantom{=} -(1-P_2)\ln(1-P_2)\nonumber \\[6pt]
& \phantom{=} + (1-P_1-P_2)\ln(1-P_1-P_2).
\end{align}
In this case we have $\max [\operatorname{I}(C_1;C_2)] = \ln 2$, which is the maximum value attainable by the mutual information of two dichotomic discrete random variables. Therefore,
\begin{align}\label{wp130}
\frac{\max [\operatorname{I}(W_1;C_2)]}{\max [\operatorname{I}(C_1;C_2)]} \approx 0.26 \sim \frac{1}{3}.
\end{align}
This result tells us that the maximum value of the mutual information of $W_1$ and $C_2$, which are linearly uncorrelated, is about one third of the maximum of the mutual information of $C_1$ and $C_2$, which can be, instead, perfectly anticorrelated. Thus, the nonlinear correlation between $W_1$ and $C_2$ is by no means negligible.

\section{Conclusions}\label{conc}

Discussions on the interpretation of light phenomena in terms of waves or particles are centuries old \cite{Hecht}. When finally the wave nature of light took over at the end of the $19^\text{th}$ century, quantum mechanics arrived to challenge it again. Nowadays, the debate on the particle versus wave interpretation of light is still ongoing and is largely based on the fact that waves interfere, while particles do not. In fact,  ``wave-particle duality'' has recently become synonymous with ``interferometric duality'' \cite{PhysRevLett.77.2154}.
The purpose of this paper was to present the wave-particle duality from a novel point of view, based on the very idea that the amplitudes of waves can vary smoothly, and  particles can be counted. Thus, we have described  wave-particle duality as continuous-discrete duality.
The two main results of this work can be summarized as follows.
\begin{enumerate}
  \item We have shown the existence of a nontrivial complementarity between the continuous and discrete nature of the electromagnetic field prepared in a single-photon state. In practice, it is not possible to simultaneously measure a nonzero wave  amplitude  not yielded by the vacuum field, and to count one photon in two separate parts of the same beam of light.
  \item We have found that the continuous and discrete random variables representing  the results of  repeated measurements of the wave amplitudes and counting of the photons, respectively, are linearly uncorrelated but not independent. We use mutual information as a statistical measure of such dependence.
\end{enumerate}
We would like to remark that our study does not cover fermionic (matter) fields.
However, the extension of our formalism to fermionic fields should be straightforward \cite{PhysRevA.58.4904,PhysRevA.88.012130}.
In the end,  we believe that this work can stimulate the use of the probabilistic techniques used here, in various applications of quantum mechanics. For example, it could be interesting to test local realism (Bell's inequality and the like), using higher-order correlation functions and mutual information to quantify the distance between classical and quantum probability distributions.

\section*{Acknowledgements}

I acknowledge support from the Deutsche Forschungsgemeinschaft Project No. 429529648-
TRR 306 QuCoLiMa (``Quantum Cooperativity of Light and Matter''). Many thanks to Valerio Scarani for useful comments on a preliminary version of this work.

\onecolumn\newpage

\appendix

\numberwithin{equation}{section}

\section{Commutation relations}\label{commrel}

\subsection{Amplitude operators}

Let $\{ f \} = \{f_1(\mathbf{x}), f_2(\mathbf{x}), \ldots , f_M(\mathbf{x}) \}$ be a set of $M$ smooth real functions, such that
\begin{align}\label{s10}
\left( f_n, f_m \right) & =   \int_{\mathbb{R}^2}   f_n(\mathbf{x}) f_m(\mathbf{x}) \, \mathrm{d} \mathbf{x} \nonumber \\[6pt]
& =   \delta_{nm}, \qquad \qquad (n,m=1,2, \ldots, M),
\end{align}
where, here and hereafter, $\mathrm{d} \mathbf{x} = \mathrm{d}x \, \mathrm{d}y$, $\mathrm{d} \mathbf{x}' = \mathrm{d}x' \, \mathrm{d}y'$, et cetera. Given the field
\begin{align}\label{s20}
\hat{\Phi} (\mathbf{x},z, t) & =   \frac{1}{\sqrt{2}}  \left[ e^{- i \omega t} \hat{\phi} (\mathbf{x},z)  +  e^{i \omega t} \hat{\phi}^\dagger (\mathbf{x},z)  \right],
\end{align}
with
\begin{align}\label{s30}
\hat{\phi} (\mathbf{x},z) & =   \sum_{\mu} \hat{a}_{\mu} u_{\mu} (\mathbf{x},z),
\end{align}
we can use the functions $\{ f \} $ to build the  set of $M$  Hermitian operators $\bigl\{ \hat{\Phi}(z) \bigr\} =$ \\ $\bigl\{\hat{\Phi}_1(z), \hat{\Phi}_2(z), \ldots , \hat{\Phi}_M(z) \bigr\}$, defined by
\begin{align}\label{s40}
\hat{W}_n(z) & =   \bigl( f_n, \,\hat{\Phi} (\mathbf{x},z,0) \bigr) \nonumber \\[6pt]
& =   \int_{\mathbb{R}^2}   f_n(\mathbf{x}) \hat{\Phi} (\mathbf{x},z,0) \, \mathrm{d} \mathbf{x} \nonumber \\[6pt]
& =  \frac{1}{\sqrt{2}}\sum_\mu \left( \hat{a}_\mu f_{n \mu}^* + \hat{a}_\mu^\dagger f_{n \mu} \right), \qquad \qquad (n=1,2, \ldots, M),
\end{align}
where $f_{n \mu} = f_{n \mu}(z)$, with
\begin{align}\label{s50}
f_{n \mu}(z) & =  \left( u_\mu, f_n \right) \nonumber \\[6pt]
& =   \int_{\mathbb{R}^2}  u_\mu^*(\mathbf{x},z)f_n(\mathbf{x}) \, \mathrm{d} \mathbf{x} .
\end{align}

Next, we calculate the commutator
\begin{align}\label{s60}
\bigl[ \hat{W}_n(z), \, \hat{W}_m(z) \bigr] & =  \bigl[ ( f_n,\hat{\Phi} ) , \, ( f_m,\hat{\Phi} ) \bigr]  \nonumber \\[6pt]
& =   \frac{1}{2} \sum_{\mu, \nu} \bigl[ {\hat{a}_\mu} ( f_n, u_\mu ) + {\hat{a}_\mu^\dagger} ( f_n, u_\mu^* ) , \, {\hat{a}_\nu} ( f_m, u_\nu ) + {\hat{a}_\nu^\dagger} ( f_m, u_\nu^* ) \bigr] \nonumber \\[6pt]
& =  \frac{1}{2} \sum_{\mu} \bigl[ {( f_n, u_\mu )( u_\mu, f_m )} - {( f_m, u_\mu )( u_\mu, f_n )} \bigr]
\nonumber \\[6pt]
& =  \frac{1}{2}  \bigl[ {( f_n,  f_m )} - {( f_m,  f_n )} \bigr]
\nonumber \\[6pt]
& =  0,
\end{align}
where we have used
\begin{align}\label{s70}
\bigl[ \hat{a}_\mu, \, \hat{a}_\nu^\dagger \bigr] & =  \delta_{\mu \nu}, \qquad \bigl[ \hat{a}_\mu, \, \hat{a}_\nu \bigr]  = 0, \qquad \text{and} \qquad \sum_{\mu} u_\mu(\mathbf{x},z)u_\mu^*(\mathbf{x}',z) = \delta(\mathbf{x} - \mathbf{x}'),
\end{align}
with $\delta(\mathbf{x} - \mathbf{x}') = \delta(x - x') \delta(y - y')$.
Note that from $f_n(\mathbf{x}) \in \mathbb{R}$ it follows that $( f_n,  f_m ) = ( f_m,  f_n )$, so that the commutator \eqref{s60} would be zero even if \eqref{s10} were not satisfied.

\subsection{Intensity operators}

Let us define the ``intensity'' operator $\hat{\mathrm{I}}(\mathbf{x},z)$ as,
\begin{align}\label{s80}
\hat{\mathrm{I}}(\mathbf{x},z) & =  \hat{\phi}^\dagger (\mathbf{x},z) \hat{\phi} (\mathbf{x},z) \nonumber \\[6pt]
& =   \sum_{\mu, \nu} \hat{a}_{\mu}^\dagger\hat{a}_{\nu}  u_{\mu}^* (\mathbf{x},z) u_{\nu} (\mathbf{x},z).
\end{align}
By definition
\begin{align}\label{s90}
\int_{\mathbb{R}^2} \hat{\mathrm{I}}(\mathbf{x},z) \, \mathrm{d} \mathbf{x}  & =   \sum_{\mu} \hat{a}_{\mu}^\dagger \hat{a}_{\mu} ,
\end{align}
where the orthogonality relation
\begin{align}\label{s100}
\left( u_\mu, \, u_\nu \right) & =  \int_{\mathbb{R}^2}  u_{\mu}^* (\mathbf{x},z) u_{\nu} (\mathbf{x},z) \,\mathrm{d} \mathbf{x}   = \delta_{\mu \nu},
\end{align}
has been used.

Let $\{ \bm{1} \} = \{\bm{1}_1(\mathbf{x}), \bm{1}_2(\mathbf{x}), \ldots , \bm{1}_M(\mathbf{x}) \}$ be a set of $M$ indicator functions with disjoint compact supports. By definition
\begin{align}\label{s110}
\operatorname{supp}\{ \bm{1}_n \} \bigcap \operatorname{supp}\{ \bm{1}_m \} & =   \varnothing \qquad \text{if} \qquad n \neq m,
\end{align}
or, equivalently,
\begin{align}\label{s120}
 \bm{1}_n(\mathbf{x}) \bm{1}_m (\mathbf{x}) & =   \delta_{nm} \bm{1}_n(\mathbf{x}),  \qquad \qquad (n,m=1,2, \ldots, M).
\end{align}
We assume that they are normalized to some areas $\mathcal{A}_n$ (typically the area of the active surface of a detector), defined by
\begin{align}\label{s130}
\int_{\mathbb{R}^2} \bm{1}_n(\mathbf{x}) \, \mathrm{d} \mathbf{x} = \int_{\mathcal{D}_n}\mathrm{d} \mathbf{x}   =   \mathcal{A}_n,
\end{align}
where $\mathcal{D}_n = \operatorname{supp}\{ \bm{1}_n \}$. Using these functions, we define the ``counting operator'' $\hat{C}_n$, as
\begin{align}\label{s140}
\hat{C}_n  & =   \bigl( \bm{1}_n, \, \hat{\mathrm{I}} (\mathbf{x},z) \bigr) \nonumber \\[6pt]
& =  \int_{\mathbb{R}^2}   \bm{1}_n(\mathbf{x}) \, \hat{\mathrm{I}}(\mathbf{x},z) \, \mathrm{d} \mathbf{x}  \nonumber \\[6pt]
& =   \sum_{\mu, \nu} \hat{a}_{\mu}^\dagger\hat{a}_{\nu} \int_{\mathbb{R}^2}    u_{\mu}^* (\mathbf{x},z) \bm{1}_n(\mathbf{x}) u_{\nu} (\mathbf{x},z)  \, \mathrm{d} \mathbf{x} \nonumber \\[6pt]
& =  \sum_{\mu, \nu} \hat{a}_\mu^\dagger \hat{a}_\nu \bm{1}_{n \mu \nu},
\end{align}
where $\bm{1}_{n \mu \nu} = \bm{1}_{n \mu \nu}(z)$, with
\begin{align}\label{s150}
\bm{1}_{n \mu \nu}(z) & =  (u_\mu, \bm{1}_n u _\nu) .
\end{align}

Next, we calculate the commutator
\begin{align}\label{s160}
\bigl[ \hat{C}_n(z), \, \hat{C}_m(z) \bigr] & =  \bigl[ ( \bm{1}_n, \hat{\mathrm{I}} ) , \, ( \bm{1}_m, \hat{\mathrm{I}} ) \bigr]  \nonumber \\[6pt]
& =   \sum_{\mu, \nu}  \sum_{\alpha, \beta} \bigl[
\hat{a}_\mu^\dagger \hat{a}_\nu \bm{1}_{n\mu \nu} , \,
\hat{a}_\alpha^\dagger \hat{a}_\beta \bm{1}_{m \alpha \beta} \bigr] \nonumber \\[6pt]
& =  \sum_{\mu, \nu}  \sum_{\alpha, \beta} \bm{1}_{n\mu \nu} \bm{1}_{m \alpha \beta} \bigl[
\hat{a}_\mu^\dagger \hat{a}_\nu , \,
\hat{a}_\alpha^\dagger \hat{a}_\beta \bigr].
\end{align}
Using the commutator relation
\begin{align}\label{s170}
\bigl[ \hat{A} \hat{B} , \, \hat{C} \hat{D} \bigr] & =  \hat{A}\bigl[\hat{B}, \hat{C}\bigr]\hat{D} + \bigl[\hat{A}, \hat{C}\bigr]\hat{B}\hat{D} + \hat{C}\hat{A}\bigl[\hat{B}, \hat{D}\bigr] + \hat{C}\bigl[\hat{A}, \hat{D}\bigr]\hat{B},
\end{align}
with $\hat{A} = \hat{a}_\mu^\dagger, \, \hat{B} = \hat{a}_\nu, \, \hat{C} = \hat{a}_\alpha^\dagger$ and $\hat{D} = \hat{a}_\beta$, we find
\begin{align}\label{s180}
\bigl[
\hat{a}_\mu^\dagger \hat{a}_\nu , \,
\hat{a}_\alpha^\dagger \hat{a}_\beta \bigr] & =  \hat{a}_\mu^\dagger \underbrace{\bigl[\hat{a}_\nu, \hat{a}_\alpha^\dagger\bigr]}_{= \, \delta_{\nu \alpha}} \hat{a}_\beta
+ \underbrace{\bigl[\hat{a}_\mu^\dagger, \hat{a}_\alpha^\dagger\bigr]}_{= \, 0}\hat{a}_\nu\hat{a}_\beta
+ \hat{a}_\alpha^\dagger\hat{a}_\mu^\dagger \underbrace{\bigl[\hat{a}_\nu, \hat{a}_\beta\bigr] }_{= \, 0}
+ \hat{a}_\alpha^\dagger \underbrace{\bigl[\hat{a}_\mu^\dagger, \hat{a}_\beta\bigr]}_{= \, - \delta_{\mu \beta}} \hat{a}_\nu   \nonumber \\[6pt]
& =  \hat{a}_\mu^\dagger \hat{a}_\beta \, \delta_{\nu \alpha} - \hat{a}_\alpha^\dagger  \hat{a}_\nu \, \delta_{\mu \beta}.
\end{align}
Substituting \eqref{s180} into \eqref{s170}, we obtain
\begin{align}\label{s190}
\bigl[ \hat{C}_n(z), \, \hat{C}_m(z) \bigr] & =  \sum_{\mu, \nu}  \sum_{\alpha, \beta} \bm{1}_{n\mu \nu} \bm{1}_{m \alpha \beta} \left( \hat{a}_\mu^\dagger \hat{a}_\beta \, \delta_{\nu \alpha} - \hat{a}_\alpha^\dagger  \hat{a}_\nu \, \delta_{\mu \beta} \right)  \nonumber \\[6pt]
& =   \sum_{\mu, \beta} \hat{a}_\mu^\dagger \hat{a}_\beta \, \sum_{\nu} \bm{1}_{n \mu \nu} \bm{1}_{m \nu \beta}  -
\sum_{\nu, \alpha} \hat{a}_\alpha^\dagger  \hat{a}_\nu \, \sum_{\mu} \bm{1}_{m \alpha \mu} \bm{1}_{n\mu \nu} .
\end{align}
Now it remains to calculate
\begin{align}\label{s200}
\sum_{\nu} \bm{1}_{n \mu \nu} \bm{1}_{m \nu \beta} & =  \sum_{\nu} (u_\mu, \bm{1}_n u _\nu) (u_\nu, \bm{1}_m u _\beta)   \nonumber \\[6pt]
& =   \sum_{\nu} \int_{\mathbb{R}^2}  u_\mu(\mathbf{x},z) \bm{1}_n (\mathbf{x} ) u _\nu(\mathbf{x},z) \,  \mathrm{d} \mathbf{x} \int_{\mathbb{R}^2} u_\nu(\mathbf{x}',z) \bm{1}_m(\mathbf{x}') u _\beta (\mathbf{x}',z) \, \mathrm{d} \mathbf{x}' \nonumber \\[6pt]
& =    \int_{\mathbb{R}^2} \Biggr\{ u_\mu(\mathbf{x},z) \bm{1}_n (\mathbf{x} ) \int_{\mathbb{R}^2}  \biggl[   \bm{1}_m(\mathbf{x}') u _\beta (\mathbf{x}',z) \underbrace{\sum_{\nu}u _\nu(\mathbf{x},z)u_\nu(\mathbf{x}',z)}_{= \, \delta(\mathbf{x} - \mathbf{x}')} \biggr] \mathrm{d} \mathbf{x}' \Biggl\} \, \mathrm{d} \mathbf{x}   \nonumber \\
& =  \int_{\mathbb{R}^2}  u_\mu(\mathbf{x},z) \bm{1}_n (\mathbf{x} )    \bm{1}_m(\mathbf{x}) u_\beta (\mathbf{x},z) \, \mathrm{d} \mathbf{x}   \nonumber \\[6pt]
& =  \bigl( u_\mu , \bm{1}_n \bm{1}_m  u_\beta \bigr),
\end{align}
where the completeness relation \eqref{s70} has been used. Finally, substituting \eqref{s200} into \eqref{s190}, we obtain
\begin{align}\label{s210}
\bigl[ \hat{C}_n(z), \, \hat{C}_m(z) \bigr] & =   \sum_{\mu, \beta} \hat{a}_\mu^\dagger \hat{a}_\beta \, \bigl( u_\mu , \bm{1}_n \bm{1}_m  u_\beta \bigr)  -
\sum_{\nu, \alpha} \hat{a}_\alpha^\dagger  \hat{a}_\nu \, \bigl( u_\alpha , \bm{1}_m \bm{1}_n  u_\nu \bigr) \nonumber \\[6pt]
& =  \sum_{\mu, \nu} \hat{a}_\mu^\dagger \hat{a}_\nu \Bigl[ \bigl( u_\mu , \bm{1}_n \bm{1}_m  u_\nu \bigr) - \bigl( u_\mu , \bm{1}_m \bm{1}_n  u_\nu \bigr)  \, \Bigr]  \nonumber \\[6pt]
& =  0,
\end{align}
where we have renamed the dummy indexes $\beta \to \nu$ and $\alpha \to \mu$. The final result comes from the trivial identity $\bm{1}_n (\mathbf{x} )    \bm{1}_m(\mathbf{x}) = \bm{1}_m (\mathbf{x} )    \bm{1}_n(\mathbf{x})$.

\subsection{A remark}

It should be noticed that actually the two commutators \eqref{s60} and \eqref{s160} are trivially equal to zero, because they are both of the form
\begin{align}\label{s220}
\bigl[ ( v, \hat{A} ) , \, ( w, \hat{A} ) \bigr] & =  \int_{\mathbb{R}^2} \left\{ \int_{\mathbb{R}^2}  v(\mathbf{x}) w(\mathbf{x}')  \bigl[ \hat{A}(\mathbf{x})  , \, \hat{A}(\mathbf{x}')  \bigr] \mathrm{d} \mathbf{x}' \right\} \mathrm{d} \mathbf{x}  ,
\end{align}
with $v(\mathbf{x}), w(\mathbf{x}') \in \mathbb{R}$.
Then, one should simply verify that
\begin{align}\label{s230}
\bigl[ \hat{\Phi}(\mathbf{x})  , \, \hat{\Phi}(\mathbf{x}')  \bigr] & =  0, \qquad \text{and} \qquad   \bigl[ \hat{\mathrm{I}}(\mathbf{x})  , \, \hat{\mathrm{I}}(\mathbf{x}')  \bigr]=0 .
\end{align}

To illustrate this procedure, we calculate now the mixed commutator
\begin{align}\label{m10}
\bigl[ \hat{W}_m(z), \, \hat{C}_n(z) \bigr] & =  \bigl[ ( f_m, \hat{\Phi} ) , \, ( \bm{1}_n, \hat{\mathrm{I}} ) \bigr]  \nonumber \\[6pt]
& =  \int_{\mathbb{R}^2} \mathrm{d} \mathbf{x} \int_{\mathbb{R}^2} \mathrm{d} \mathbf{x}' \, f_m(\mathbf{x}) \bm{1}_n(\mathbf{x}')  \bigl[ \hat{\Phi}(\mathbf{x},z,0)  , \, \hat{\mathrm{I}}(\mathbf{x}',z)  \bigr],
\end{align}
where
\begin{align}\label{m20}
\bigl[ \hat{\Phi}(\mathbf{x},z,0)  , \, \hat{\mathrm{I}}(\mathbf{x}',z)  \bigr] & =  \frac{1}{\sqrt{2}} \bigl[ \hat{\phi} (\mathbf{x},z)  +   \hat{\phi}^\dagger (\mathbf{x},z) , \, \hat{\phi}^\dagger (\mathbf{x}',z)  \hat{\phi} (\mathbf{x}',z) \bigr]  \nonumber  \\[6pt]
& =  \frac{1}{\sqrt{2}} \left\{ \bigl[ \hat{\phi} (\mathbf{x},z)   , \, \hat{\phi}^\dagger (\mathbf{x}',z)  \hat{\phi} (\mathbf{x}',z) \bigr] + \bigl[  \hat{\phi}^\dagger (\mathbf{x},z) , \, \hat{\phi}^\dagger (\mathbf{x}',z)  \hat{\phi} (\mathbf{x}',z) \bigr] \right\} .
\end{align}
Using
\begin{align}\label{m30}
\bigl[ \hat{A}  , \, \hat{B} \hat{C}  \bigr] & =  \bigl[ \hat{A}  , \, \hat{B} \bigr] \hat{C} + \hat{B} \bigl[ \hat{A} , \, \hat{C}  \bigr],
\end{align}
and
\begin{align}\label{m40}
\left[ \hat{\phi} (\mathbf{x},z)  , \, \hat{\phi}^\dagger (\mathbf{x}',z)  \right] & =  \Bigl[  \sum_{\mu} \hat{a}_{\mu} u_{\mu} (\mathbf{x},z)  , \,  \sum_{\nu} \hat{a}_{\nu}^\dagger u_{\nu}^* (\mathbf{x}',z) \Bigr] \nonumber  \\[6pt]
& =   \sum_{\mu, \nu}u_{\mu} (\mathbf{x},z) u_{\nu}^* (\mathbf{x}',z) \underbrace{\Bigl[  \hat{a}_{\mu} , \,  \hat{a}_{\nu}^\dagger \Bigr]}_{= \, \delta_{\mu \nu}} \nonumber  \\[6pt]
& =   \sum_{\mu}u_{\mu} (\mathbf{x},z) u_{\mu}^* (\mathbf{x}',z)  \nonumber  \\[6pt]
& =   \delta \left( \mathbf{x} - \mathbf{x}' \right),
\end{align}
we calculate straightforwardly
\begin{align}\label{m50}
\Bigl[ \hat{\phi} (\mathbf{x},z)   , \, \hat{\phi}^\dagger (\mathbf{x}',z)  \hat{\phi} (\mathbf{x}',z) \Bigr] & =  \Bigl[ \hat{\phi} (\mathbf{x},z)   , \, \hat{\phi}^\dagger (\mathbf{x}',z)  \Bigr] \hat{\phi} (\mathbf{x}',z) \nonumber  \\[6pt]
& =   \delta \left( \mathbf{x} - \mathbf{x}' \right) \hat{\phi} (\mathbf{x}',z),
\end{align}
and
\begin{align}\label{m60}
\Bigl[  \hat{\phi}^\dagger (\mathbf{x},z) , \, \hat{\phi}^\dagger (\mathbf{x}',z)  \hat{\phi} (\mathbf{x}',z) \Bigr] & =  \hat{\phi}^\dagger (\mathbf{x}',z)  \Bigl[  \hat{\phi}^\dagger (\mathbf{x},z) , \, \hat{\phi} (\mathbf{x}',z) \Bigr]  \nonumber  \\[6pt]
& =  -\delta \left( \mathbf{x} - \mathbf{x}' \right) \hat{\phi}^\dagger (\mathbf{x}',z) .
\end{align}
Substituting \eqref{m50} and \eqref{m60} into \eqref{m20}, we obtain
\begin{align}\label{m70}
\bigl[ \hat{\Phi}(\mathbf{x},z,0)  , \, \hat{\mathrm{I}}(\mathbf{x}',z)  \bigr] & =  \frac{1}{\sqrt{2}} \left\{ \bigl[ \hat{\phi} (\mathbf{x},z)   , \, \hat{\phi}^\dagger (\mathbf{x}',z)  \hat{\phi} (\mathbf{x}',z) \bigr] + \bigl[  \hat{\phi}^\dagger (\mathbf{x},z) , \, \hat{\phi}^\dagger (\mathbf{x}',z)  \hat{\phi} (\mathbf{x}',z) \bigr] \right\}  \nonumber  \\[6pt]
& =  \frac{1}{\sqrt{2}} \, \delta \left( \mathbf{x} - \mathbf{x}' \right) \left\{\hat{\phi} (\mathbf{x}',z) - \hat{\phi}^\dagger (\mathbf{x}',z) \right\} .
\end{align}
Substitution of \eqref{m70} into \eqref{m10}, gives
\begin{align}\label{m80}
\bigl[ \hat{W}_m(z), \, \hat{C}_n(z) \bigr] & =  \int_{\mathbb{R}^2} \frac{1}{\sqrt{2}} \, f_m(\mathbf{x})\left\{ \int_{\mathbb{R}^2}  \bm{1}_n(\mathbf{x}')  \, \delta \left( \mathbf{x} - \mathbf{x}' \right) \left[\hat{\phi} (\mathbf{x}',z) - \hat{\phi}^\dagger (\mathbf{x}',z) \right] \mathrm{d} \mathbf{x}'  \right\}  \mathrm{d} \mathbf{x} \nonumber \\[6pt]
& =  \int_{\mathbb{R}^2}  f_m(\mathbf{x}) \bm{1}_n(\mathbf{x})  \frac{\hat{\phi} (\mathbf{x},z) - \hat{\phi}^\dagger (\mathbf{x},z)}{\sqrt{2}} \, \mathrm{d} \mathbf{x}\nonumber \\[6pt]
& =  \int_{\mathcal{D}_n}  f_m(\mathbf{x})  \frac{\hat{\phi} (\mathbf{x},z) - \hat{\phi}^\dagger (\mathbf{x},z)}{\sqrt{2}} \, \mathrm{d} \mathbf{x}\nonumber \\[6pt]
& =  \delta_{nm}\int_{\mathcal{D}_n}  f_n(\mathbf{x})  \frac{\hat{\phi} (\mathbf{x},z) - \hat{\phi}^\dagger (\mathbf{x},z)}{\sqrt{2}} \, \mathrm{d} \mathbf{x} ,
\end{align}
because, by hypothesis,  $ f_m(\mathbf{x}) \bm{1}_n(\mathbf{x}) = 0$ for  $n \neq m$. Then, we can rewrite \eqref{m80} in a more compact and suggestive form as,
\begin{align}\label{m85}
\bigl[ \hat{W}_m(z), \, \hat{C}_n(z) \bigr] = \frac{\delta_{nm}}{\sqrt{2}} \left\{ \bigl(f_n, \hat{\phi}  \bigr) - \bigl(\hat{\phi}, f_n \bigr) \right\}.
\end{align}

\section{Probability distribution for the wave operators}\label{pdfW}

Here we calculate step by step the following expression for $p_\mathbf{W} (N,\mathbf{w})$, which is defined by
\begin{align}\label{s240}
p_\mathbf{W}(N,\mathbf{w})  = \langle N[\phi] | \prod_{n = 1}^M \delta \bigl( \hat{W}_n - {w}_n \bigr) | N[\phi] \rangle,
\end{align}
for $N=0$ and $N=1$. In the remainder we will use  the Fourier transform representation of the Dirac delta function,
\begin{align}\label{s250}
\delta(x-x_0) & =   \frac{1}{2 \pi} \int_{\mathbb{R}} e^{i \alpha (x - x_0)} \, \mathrm{d} \alpha .
\end{align}

\subsection{Vacuum state}

\begin{align}\label{s260}
p_\mathbf{W}(0,\mathbf{w}) & =  \langle 0 | \prod_{n = 1}^M \delta \bigl( \hat{W}_n - w_n \bigr) | 0 \rangle \nonumber \\[6pt]
& =  \frac{1}{(2 \pi)^M} \int_{\mathbb{R}} \mathrm{d} \alpha_1 \cdots \int_{\mathbb{R}} \mathrm{d} \alpha_M \, \exp\left( - i \sum_{n=1}^M \alpha_n w_n \right) \langle 0 | e^{i \alpha_1 \hat{W}_1 } e^{i \alpha_2 \hat{W}_2 } \cdots e^{i \alpha_M \hat{W}_M }| 0 \rangle \nonumber \\[6pt]
& =  \frac{1}{(2 \pi)^M} \int_{\mathbb{R}} \mathrm{d} \alpha_1 \cdots \int_{\mathbb{R}} \mathrm{d} \alpha_M \, \exp\left( - i \sum_{n=1}^M \alpha_n w_n \right) \langle 0 | \exp\left(  i \sum_{n=1}^M \alpha_n \hat{W}_n \right) | 0 \rangle,
\end{align}
where \eqref{s60} has been used. Next, using \eqref{s40} and \eqref{s50}, we rewrite
\begin{align}\label{s270}
\sum_{n=1}^M \alpha_n \hat{W}_n  & =  \sum_\mu \left( \hat{a}_\mu \varphi_{\mu}^* + \hat{a}_\mu^\dagger \varphi_{\mu} \right)   ,
\end{align}
where using \eqref{s50} we have defined
\begin{align}\label{s280}
\varphi_{\mu} & =  \frac{1}{\sqrt{2}} \sum_{n=1}^M \alpha_n  f_{n \mu} \nonumber \\[6pt]
& =  \biggl( u_\mu , \frac{1}{\sqrt{2}} \sum_{n=1}^M \alpha_n  f_{n } \biggr)\nonumber \\[6pt]
& =  \bigl( u_\mu , \varphi \bigr).
\end{align}
This implies that we can define the field $\varphi(\mathbf{x})$ as,
\begin{align}\label{s290}
\varphi(\mathbf{x}) & =   \frac{1}{\sqrt{2}} \sum_{n=1}^M \alpha_n  f_{n }(\mathbf{x}).
\end{align}

Now, we are ready to calculate
\begin{align}\label{s300}
\langle 0 | \exp\left(  i \sum_{n=1}^M \alpha_n \hat{W}_n \right) | 0 \rangle & =  \langle 0 | e^{\hat{A} + \hat{B}} | 0 \rangle,
\end{align}
where we have defined
\begin{align}\label{s310}
\hat{A} & =    i \sum_\mu \hat{a}_\mu \varphi_{\mu}^* , \qquad \text{and } \qquad \hat{B}  =    i \sum_\mu \hat{a}_\mu^\dagger \varphi_{\mu}.
\end{align}
It is easy to see that
\begin{align}\label{s320}
\bigl[ \hat{A}, \, \hat{B} \bigr] & =  - \sum_{\mu, \nu} \varphi_{\mu}^* \varphi_{\nu} \bigl[ \hat{a}_\mu, \,  \hat{a}_\nu^\dagger \bigr] \nonumber \\[6pt]
& =  - \sum_{\mu} | \varphi_{\mu} |^2  \nonumber \\[6pt]
& =  - \bigl( \varphi, \varphi \bigr) \nonumber \\[6pt]
& =  - \frac{1}{2} \sum_{n,m=1}^M \alpha_n \alpha_m \bigl(f_n, f_m \bigr) \nonumber \\[6pt]
& =  - \frac{1}{2} \sum_{n=1}^M \alpha_n^2 \bigl(f_n, f_n \bigr) ,
\end{align}
where \eqref{s290} and \eqref{s10}  have been used.
So, we can use the Campbell-Baker-Hausdorff identity \cite{MandelBook},
\begin{align}\label{s330}
e^{ \hat{A} + \hat{B} } = e^{ \hat{A}} \, e^{ \hat{B}  }  \, e^{-\frac{1}{2}[\hat{A},\hat{B}]} = e^{ \hat{B}} \, e^{ \hat{A}  }  \, e^{\frac{1}{2}[\hat{A},\hat{B}]}, \qquad \text{if} \qquad [\hat{A},[\hat{A},\hat{B}]] = 0 = [\hat{B},[\hat{A},\hat{B}]],
\end{align}
to rewrite
\begin{align}\label{s340}
\langle 0 | \exp\left(  i \sum_{n=1}^M \alpha_n \hat{W}_n \right) | 0 \rangle & =  \langle 0 | e^{\hat{A} + \hat{B}} | 0 \rangle \nonumber \\[6pt]
& =  e^{\frac{1}{2}[\hat{A},\hat{B}]} \underbrace{\langle 0 |  e^{ \hat{B}} \, e^{ \hat{A}  }  | 0 \rangle}_{= \, 1} \nonumber \\[6pt]
& =  \exp \left[ - \frac{1}{4} \sum_{n=1}^M \alpha_n^2 \bigl(f_n, f_n \bigr) \right],
\end{align}
where we have used $\hat{A} | 0 \rangle = 0 = \langle 0 | \hat{B}$. Substituting \eqref{s340} into \eqref{s260} we obtain
\begin{align}\label{s350}
p_\mathbf{W} (0,\mathbf{w}) & =  \frac{1}{(2 \pi)^M} \int_{\mathbb{R}} \mathrm{d} \alpha_1 \cdots \int_{\mathbb{R}} \mathrm{d} \alpha_M \, \exp\left( - i \sum_{n=1}^M \alpha_n w_n \right) \left[ - \frac{1}{4} \sum_{n=1}^M \alpha_n^2 \bigl(f_n, f_n \bigr) \right] \nonumber \\[6pt]
& =  \prod_{n=1}^M   \int_{\mathbb{R}} \frac{\mathrm{d} \alpha_n}{2 \pi}  \exp\left[ - \frac{\bigl(f_n, f_n \bigr)}{4} \alpha_n^2 - i w_n  \alpha_n \right] \nonumber \\[6pt]
& =  \prod_{n=1}^M  \frac{1}{\sqrt{\pi \bigl(f_n, f_n \bigr)}} \exp \left[ - \frac{w_n^2}{\bigl(f_n, f_n \bigr)} \right]
\nonumber \\[6pt]
& \deff \prod_{n=1}^M p_{W_n}(0,w_n),
\end{align}
where the following Gaussian integral has been used (see Eq. (3.16) in \cite{SchulmanBook}):
\begin{align}\label{s360}
\int_{-\infty}^{\infty} \exp\left( -a y^2 + b y \right ) \mathrm{d} y  = \sqrt{ \frac{\pi}{a} } \exp\left( \frac{b^2}{4 a}\right ),
\end{align}
with $a,b \in \mathbb{C}$, and $\text{Re} \, a > 0$. Note that in the main text  we  use  the shortcut
\begin{align}\label{s365}
p_{W}(0,w) = \frac{1}{\sqrt{2 \pi} \, \sigma} \exp \left( - \frac{w^2}{2 \sigma^2} \right) \deff p_0(w) ,
\end{align}
with $\sigma^2 = (f,f)/2$, when $M=1$, and
\begin{align}\label{s366}
p_{W_1 W_2}(0,w_1,w_2) =  p_0(w_1) p_0(w_2),
\end{align}
when $M=2$.

\subsection{Single photon state}

In this case we calculate the probability distribution with respect to the single-photon state $| 1[\phi] \rangle $ defined by
\begin{align}\label{s370}
| 1[\phi] \rangle  & =  \hat{a}^\dagger[\phi] |0 \rangle \nonumber \\[6pt]
& =   \sum_\mu \phi_\mu \hat{a}^\dagger_\mu |0 \rangle,
\end{align}
where, by hypothesis,
\begin{align}\label{s380}
\phi_\mu & =  \bigl(u_\mu, \phi \bigr),  \qquad \text{with} \qquad \bigl(\phi, \phi \bigr) = 1.
\end{align}
So, we must evaluate
\begin{align}\label{s390}
p_\mathbf{W} (1,\mathbf{w}) & =  \langle 1[\phi] | \prod_{n = 1}^M \delta \bigl( \hat{W}_n - w_n \bigr) | 1[\phi] \rangle \nonumber \\[6pt]
& =  \sum_{\mu, \nu} \phi_\mu^* \phi_\nu \langle 0 | \hat{a}_\mu \prod_{n = 1}^M \delta \bigl( \hat{W}_n - w_n \bigr) \hat{a}_\nu^\dagger | 0 \rangle  ,
\end{align}
where
\begin{multline}\label{s400}
\langle 0 | \hat{a}_\mu \prod_{n = 1}^M \delta \bigl( \hat{W}_n - w_n \bigr) \hat{a}_\nu^\dagger | 0 \rangle \\[6pt]
= \frac{1}{(2 \pi)^M} \int_{\mathbb{R}} \mathrm{d} \alpha_1 \cdots \int_{\mathbb{R}} \mathrm{d} \alpha_M \, \exp\left( - i \sum_{n=1}^M \alpha_n w_n \right)  \langle 0 | \hat{a}_\mu \exp\left(  i \sum_{n=1}^M \alpha_n \hat{W}_n \right) \hat{a}_\nu^\dagger| 0 \rangle.
\end{multline}
Proceeding like in the previous section and using \eqref{s310}, we can write
\begin{align}\label{s410}
\langle 0 | \hat{a}_\mu \exp\left(  i \sum_{n=1}^M \alpha_n \hat{W}_n \right) \hat{a}_\nu^\dagger| 0 \rangle & =  \langle 0 |\hat{a}_\mu   e^{\hat{A} + \hat{B}} \hat{a}_\nu^\dagger | 0 \rangle \nonumber \\[6pt]
& =  e^{\frac{1}{2}[\hat{A},\hat{B}]} \langle 0 |\hat{a}_\mu e^{ \hat{B}} \, e^{ \hat{A}  }  \hat{a}_\nu^\dagger | 0 \rangle \nonumber \\[6pt]
& =  e^{\frac{1}{2}[\hat{A},\hat{B}]} \langle 0 |e^{ \hat{B}}  \left( e^{ -\hat{B}} \hat{a}_\mu e^{ \hat{B}} \right) \left( e^{ \hat{A}  }  \hat{a}_\nu^\dagger e^{ -\hat{A}} \right)e^{ \hat{A}} | 0 \rangle \nonumber \\[6pt]
& =  e^{\frac{1}{2}[\hat{A},\hat{B}]} \langle 0 | \left(  \hat{a}_\mu - \bigl[ \hat{B} , \, \hat{a}_\mu \bigr]  \right) \left(  \hat{a}_\nu^\dagger + \bigl[ \hat{A} , \, \hat{a}_\nu^\dagger \bigr]  \right) | 0 \rangle \nonumber \\[6pt]
& =  e^{\frac{1}{2}[\hat{A},\hat{B}]} \left\{ \langle 0 |  \hat{a}_\mu  \hat{a}_\nu^\dagger | 0 \rangle - \bigl[ \hat{B} , \, \hat{a}_\mu \bigr]   \bigl[ \hat{A} , \, \hat{a}_\nu^\dagger \bigr] \right\},
\end{align}
where we have used equation (10.11-2) in \cite{MandelBook}:
\begin{align}\label{s420}
\exp \bigl( x \hat{A} \bigr) \hat{B} \exp \bigl( -x \hat{A} \bigr) & =  \hat{B} + x \bigl[ \hat{A} , \, \hat{B} \bigr], \qquad \text{if} \;\bigl[ \hat{A} , \, \hat{B} \bigr] \; \text{is a $c$-number}.
\end{align}
In our case
\begin{align}\label{s430}
\bigl[ \hat{A} , \, \hat{a}_\nu^\dagger \bigr] & =   i \sum_\mu \varphi_{\mu}^* \bigl[ \hat{a}_\mu  , \, \hat{a}_\nu^\dagger \bigr] \nonumber \\[6pt]
& =   i \, \varphi_{\nu}^*,
\end{align}
and
\begin{align}\label{s440}
\bigl[ \hat{B} , \, \hat{a}_\mu \bigr] & =    i \sum_\nu  \varphi_{\nu} \bigl[ \hat{a}_\nu^\dagger , \, \hat{a}_\mu   \bigr] \nonumber \\[6pt]
& =   -i \, \varphi_{\mu}.
\end{align}
Substituting \eqref{s430} and \eqref{s440} into \eqref{s410}, we obtain
\begin{align}\label{s450}
\langle 0 | \hat{a}_\mu \exp\left(  i \sum_{n=1}^M \alpha_n \hat{W}_n \right) \hat{a}_\nu^\dagger| 0 \rangle & =  e^{\frac{1}{2}[\hat{A},\hat{B}]} \left\{ \langle 0 |  \hat{a}_\mu  \hat{a}_\nu^\dagger | 0 \rangle - \bigl[ \hat{B} , \, \hat{a}_\mu \bigr]   \bigl[ \hat{A} , \, \hat{a}_\nu^\dagger \bigr] \right\} \nonumber \\[6pt]
& =  e^{\frac{1}{2}[\hat{A},\hat{B}]} \bigl\{ \delta_{\mu \nu} - \left( - i \, \varphi_{\mu} \right) \left(  i \, \varphi_{\nu}^* \right) \bigr\}   \nonumber \\[6pt]
& =    \exp \left[ - \frac{1}{4} \sum_{n=1}^M \alpha_n^2 \bigl(f_n, f_n \bigr) \right] \bigl( \delta_{\mu \nu} - \varphi_{\mu}  \varphi_{\nu}^*  \bigr) .
\end{align}
Inserting this expression into \eqref{s390} and using \eqref{s400}, we find
\begin{align}\label{s460}
p_\mathbf{W} & (1,\mathbf{w})  \nonumber \\[6pt]
& =  \sum_{\mu, \nu} \phi_\mu^* \phi_\nu \, \langle 0 | \hat{a}_\mu \prod_{l = 1}^M \delta \bigl( \hat{W}_l - w_l \bigr) \hat{a}_\nu^\dagger | 0 \rangle \nonumber \\[6pt]
& =      \frac{1}{(2 \pi)^M} \int_{\mathbb{R}} \mathrm{d} \alpha_1 \cdots \int_{\mathbb{R}} \mathrm{d} \alpha_M \Biggl[  \exp \left\{ - \sum_{l=1}^M \left[\frac{1}{4} \alpha_l^2 \bigl(f_l, f_l \bigr) + i \alpha_l w_l  \right] \right\} \underbrace{\sum_{\mu, \nu} \phi_\mu^* \phi_\nu \bigl( \delta_{\mu \nu} - \varphi_{\mu}  \varphi_{\nu}^*  \bigr)}_{= \, ( \phi, \phi ) - |( \phi, \varphi )|^2} \Biggr] \nonumber  \\[6pt]
& =  \underbrace{\bigl( \phi, \phi \bigr)}_{= \; 1} p_\mathbf{W}(0,\mathbf{w}) \nonumber \\[6pt]
& \phantom{= .} - \frac{1}{(2 \pi)^M} \int_{\mathbb{R}} \mathrm{d} \alpha_1 \cdots \int_{\mathbb{R}} \mathrm{d} \alpha_M \, |( \phi, \varphi )|^2 \exp \left\{ - \sum_{l=1}^M \left[\frac{1}{4} \alpha_l^2 \bigl(f_l, f_l \bigr) + i \alpha_l w_l  \right] \right\}    ,
\end{align}
where
\begin{align}\label{s470}
|( \phi, \varphi )|^2 & =   \frac{1}{2} \sum_{n,m=1}^M \alpha_n \alpha_m \bigl( \phi, f_n \bigr) \bigl( f_m, \phi \bigr) ,
\end{align}
because of  \eqref{s290}. For the calculation of the multiple integral in \eqref{s460} it is useful to separate the terms with $n = m$ in \eqref{s470} from the rest of the sum, using the trivial identity
\begin{align}\label{s475}
\left(\sum_{n=1}^M a_n \right) \left(\sum_{m=1}^M b_m \right) & =   \sum_{n=1}^M a_n b_n +   \sum_{n=1}^M \sum_{m \neq n}  a_n b_m .
\end{align}
Applying this formula to \eqref{s470}, we find
\begin{align}\label{s480}
|( \phi, \varphi )|^2 & =   \frac{1}{2} \sum_{n=1}^M \alpha_n^2 \,  |\bigl( \phi, f_n \bigr)|^2 +  \frac{1}{2} \sum_{n=1}^M \sum_{m \neq n}  \alpha_n \alpha_m \bigl( \phi, f_n \bigr) \bigl( f_m, \phi \bigr) .
\end{align}
Substituting \eqref{s480} into \eqref{s460}, we obtain
\begin{align}\label{s490}
p_\mathbf{W} (1,\mathbf{w}) & =     p_\mathbf{W}(0,\mathbf{w}) -  \frac{1}{2} \sum_{n=1}^M |\bigl( \phi, f_n \bigr)|^2    \int_{\mathbb{R}} \frac{\mathrm{d} y }{2 \pi} \, y^2 \,  e^{-   y^2 \frac{(f_n, f_n ) }{4} - i y w_n }\prod_{\substack{l=1 \\ l \neq n }}^M \underbrace{\int_{\mathbb{R}} \frac{\mathrm{d} \alpha_l}{2 \pi} \, e^{-   \alpha_l^2 \frac{(f_l, f_l )}{4} - i \alpha_l w_l }}_{= \, p_{W_l}(0,w_l)}
\nonumber \\[6pt]
& \phantom{=} - \frac{1}{2} \sum_{n=1}^M \sum_{m \neq n}  \bigl( \phi, f_n \bigr) \bigl( f_m, \phi \bigr) \int_{\mathbb{R}} \frac{\mathrm{d} x }{2 \pi} \, x \, e^{-   x^2 \frac{(f_n, f_n )}{4} - i x w_n }  \int_{\mathbb{R}} \frac{\mathrm{d} y }{2 \pi} \, y \, e^{-   y^2 \frac{(f_m, f_m )}{4} - i y w_m } \nonumber \\[6pt]
& \phantom{=} \times \prod_{\substack{l=1 \\ l \neq n, m }}^M \underbrace{ \int_{\mathbb{R}} \frac{\mathrm{d} \alpha_l}{2 \pi} \, e^{- \alpha_l^2 \frac{(f_l, f_l )}{4} - i \alpha_l w_l } }_{= \, p_{W_l}(0,w_l)}  ,
\end{align}
where \eqref{s350} has been applied. Using the Gaussian integrals  Eqs. (3.17-3.18) in \cite{SchulmanBook}, namely,
\begin{align}\label{s500}
\int_{-\infty}^{\infty} y \, \exp\left( -a y^2 + b y \right ) \mathrm{d} y  & =  \frac{b}{2a} \,  \sqrt{ \frac{\pi}{a} } \exp\left( \frac{b^2}{4 a}\right ),
\end{align}
and
\begin{align}\label{s510}
\int_{-\infty}^{\infty} y^2 \, \exp\left( -a y^2 + b y \right ) \mathrm{d} y   & =  \frac{1}{2a} \, \left( 1 + \frac{b^2}{2a}\right) \,  \sqrt{ \frac{\pi}{a} } \exp\left( \frac{b^2}{4 a}\right ),
\end{align}
respectively, where $a,b \in \mathbb{C}$, with $\text{Re} \, a > 0$, we rewrite
\begin{align}\label{s495}
p_\mathbf{W} (1,\mathbf{w}) & =     p_\mathbf{W}(0,\mathbf{w}) -  \frac{1}{2} \sum_{n=1}^M |\bigl( \phi, f_n \bigr)|^2   p_{W_n}(0,w_n) \, \frac{2}{\bigl(f_n,f_n \bigr)} \left[ 1 - \frac{2 w^2_n}{\bigl(f_n,f_n \bigr)}\right] \prod_{\substack{l=1 \\ l \neq n }}^M p_{W_l}(0,w_l)
\nonumber \\[6pt]
&- \frac{1}{2} \sum_{n=1}^M \sum_{m \neq n}  \bigl( \phi, f_n \bigr) \bigl( f_m, \phi \bigr) p_{W_n}(0,w_n) \,\frac{-2 i w_n}{\bigl(f_n,f_n \bigr)}  \, p_{W_m}(0,w_m) \frac{-2 i w_m}{\bigl(f_m,f_m \bigr)} \prod_{\substack{l=1 \\ l \neq n, m }}^M p_{W_l}(0,w_l) \nonumber \\[6pt]
& =    p_\mathbf{W}(0,\mathbf{w}) \left\{ 1 -   \sum_{n=1}^M \frac{|\bigl( \phi, f_n \bigr)|^2}{\bigl(f_n,f_n \bigr)}    \left[ 1 - \frac{2 w^2_n}{\bigl(f_n,f_n \bigr)}\right] \right. \nonumber \\[6pt]
& \left. \phantom{=}
+ 2  \sum_{n=1}^M \sum_{m \neq n} w_n w_m   \frac{\bigl( \phi, f_n \bigr) }{\bigl(f_n,f_n \bigr)}   \frac{ \bigl( f_m, \phi \bigr) }{\bigl(f_m,f_m \bigr)} \right\}.
\end{align}
Next,  we define
\begin{align}\label{s520}
\sigma_n^2 = \frac{1}{2} (f_n,f_n), \qquad    \text{and} \qquad s_n = \frac{(\phi,f_n)}{(f_n,f_n)^{1/2}}  ,
\end{align}
to rewrite \eqref{s490} as
\begin{align}\label{s530}
p_\mathbf{W} (1,\mathbf{w}) & =  p_\mathbf{W}(0,\mathbf{w}) \left[ 1 -   \sum_{n=1}^M |s_n|^2 \left( 1 - \frac{ w^2_n}{\sigma_n^2}\right) +   \sum_{n=1}^M \sum_{m \neq n} \frac{w_n}{\sigma_n} \, \frac{w_m}{\sigma_m}  \,  s_n s_m^*  \right] \nonumber \\[6pt]
& =    p_\mathbf{W}(0,\mathbf{w}) \left[ 1 -   \sum_{n=1}^M |s_n|^2 + \left| \sum_{n=1}^M \, \frac{w_n}{\sigma_n} \, s_n \right|^2 \right],
\end{align}
where we have used \eqref{s475} backwards to reconstruct the modulus square of the sum.

We can rewrite \eqref{s530} in a more enlightening form introducing the $M$-component complex vector $\mathbf{s}$, defined by
\begin{align}\label{s600}
\mathbf{s} = (s_1, \ldots, s_M) \qquad \Rightarrow \qquad |\mathbf{s}|^2 = (\mathbf{s},\mathbf{s}) = |s_1|^2 + \ldots |s_M|^2.
\end{align}
From this definition it follows that
\begin{align}\label{q610}
p_\mathbf{W} (1,\mathbf{w})  & =  \bigl( 1 - |\mathbf{s}|^2 \bigr)p_\mathbf{W}(0,\mathbf{w})  +   |\mathbf{s}|^2 p_\mathbf{W}(0,\mathbf{w}) \left| \sum_{n=1}^M \, \frac{w_n}{\sigma_n} \, \frac{s_n}{|\mathbf{s}|} \right|^2 \nonumber \\[6pt]
& \deff \bigl( 1 - |\mathbf{s}|^2 \bigr) p_0(\mathbf{w})  +   |\mathbf{s}|^2 p_1(\mathbf{w}),
\end{align}
where we have re-defined
\begin{align}\label{q620}
p_0(\mathbf{w}) \deff p_\mathbf{W}(0,\mathbf{w}), \qquad \text{and} \qquad p_1(\mathbf{w}) \deff  p_\mathbf{W}(0,\mathbf{w}) \left| \sum_{n=1}^M \, \frac{w_n}{\sigma_n} \, \frac{s_n}{|\mathbf{s}|} \right|^2.
\end{align}
Note that by definition
\begin{align}\label{q630}
\int_{\mathbb{R}^M} p_0(\mathbf{w}) \, \text{d}\mathbf{w} = 1 = \int_{\mathbb{R}^M} p_1(\mathbf{w}) \, \text{d} \mathbf{w}.
\end{align}

When $M=1$ \eqref{q610} reduces to
\begin{align}\label{q640}
p_W (1,w)  = \bigl( 1 - |s|^2 \bigr) p_0(w)  +   |s|^2 p_1(w) ,
\end{align}
where $ p_0(w)$ is defined by \eqref{s365}, and
\begin{align}\label{q650}
 p_1(w) = p_0(w) \, \frac{w^2}{\sigma^2}.
\end{align}

\section{Probability distribution for the particle operators}\label{pdfC}

Here we calculate step by step the following expression for $p_\mathbf{C}(N ,\mathbf{c})$:
\begin{align}\label{s540}
p_\mathbf{C}(N ,\mathbf{c})& =  \langle N[\phi] | \prod_{n = 1}^M \delta \bigl( \hat{C}_n - {c}_n \bigr) | N[\phi] \rangle.
\end{align}

\subsection{Vacuum state}

In this case we have
\begin{align}\label{s550}
p_\mathbf{C}(0,\mathbf{c}) & =  \langle 0 | \prod_{n = 1}^M \delta \bigl( \hat{C}_n - c_n \bigr) | 0 \rangle \nonumber \\[6pt]
& =  \frac{1}{(2 \pi)^M} \int_{\mathbb{R}} \mathrm{d} \alpha_1 \cdots \int_{\mathbb{R}} \mathrm{d} \alpha_M \, \exp\left( - i \sum_{n=1}^M \alpha_n c_n \right) \langle 0 | e^{i \alpha_1 \hat{C}_1 } e^{i \alpha_2 \hat{C}_2 } \cdots e^{i \alpha_M \hat{C}_M }| 0 \rangle \nonumber \\[6pt]
& =  \frac{1}{(2 \pi)^M} \int_{\mathbb{R}} \mathrm{d} \alpha_1 \cdots \int_{\mathbb{R}} \mathrm{d} \alpha_M \, \exp\left( - i \sum_{n=1}^M \alpha_n c_n \right) \underbrace{\langle 0 | \exp\left(  i \sum_{n=1}^M \alpha_n \hat{C}_n \right) | 0 \rangle}_{= \, 1}\nonumber \\
& =  \prod_{n=1}^M \delta(c_n),
\end{align}
because $\hat{C}_n  | 0 \rangle = 0$, and  \eqref{s210} has been used. When $M=1$ we write in compact manner
\begin{align}\label{s552}
p_C(0,c)  =  \delta(c) \deff p_0(c) ,
\end{align}
consistently with the notation used for $p_0(w)$ in \eqref{s635}.

\subsection{Single photon state}

In this case, we must evaluate
\begin{align}\label{s555}
p_\mathbf{C} (1,\mathbf{c}) & =  \langle 1[\phi] | \prod_{n = 1}^M \delta \bigl( \hat{C}_n - c_n \bigr) | 1[\phi] \rangle \nonumber \\[6pt]
& =  \sum_{\mu, \nu} \phi_\mu^* \phi_\nu \langle 0 | \hat{a}_\mu \prod_{n = 1}^M \delta \bigl( \hat{C}_n - c_n \bigr) \hat{a}_\nu^\dagger | 0 \rangle  ,
\end{align}
where
\begin{multline}\label{s557}
\langle 0 | \hat{a}_\mu \prod_{n = 1}^M \delta \bigl( \hat{C}_n - c_n \bigr) \hat{a}_\nu^\dagger | 0 \rangle \\[6pt]
= \int_{\mathbb{R}} \frac{\mathrm{d}  \alpha_1}{2 \pi} \cdots \int_{\mathbb{R}} \frac{\mathrm{d}  \alpha_M}{2 \pi} \, \exp\left( - i \sum_{n=1}^M \alpha_n c_n \right)  \langle 0 | \hat{a}_\mu \exp\left(  i \sum_{n=1}^M \alpha_n \hat{C}_n \right) \hat{a}_\nu^\dagger| 0 \rangle.
\end{multline}
We need to calculate explicitly the last term of the previous equation, that is
\begin{align}\label{c10}
\langle 0 | \hat{a}_\mu \exp\left(  i \sum_{n=1}^M \alpha_n \hat{C}_n \right) \hat{a}_\nu^\dagger| 0 \rangle & = \langle 0 | \hat{a}_\mu \, e^{x \hat{A}} \hat{B}| 0 \rangle \nonumber \\[6pt]
& = \langle 0 | \hat{a}_\mu \, \left( e^{x \hat{A}} \hat{B} \, e^{-x \hat{A}}\right) e^{x \hat{A}}| 0  \rangle \nonumber \\[6pt]
& = \langle 0 | \hat{a}_\mu \, \left( e^{x \hat{A}} \hat{B} \, e^{-x \hat{A}}\right)| 0  \rangle ,
\end{align}
where we have defined
\begin{align}\label{c20}
x = i, \qquad \hat{A} = \sum_{n=1}^M \alpha_n \hat{C}_n , \qquad \text{and} \qquad \hat{B} = \hat{a}_\nu^\dagger.
\end{align}
Moreover, we have used \eqref{s140} to calculate  $\hat{C}_n | 0 \rangle = 0$, which implies $\exp ( {x \hat{A}} )| 0 \rangle = | 0 \rangle$.

Next, to calculate \eqref{c10} we need to use equation (10.11-1) in \cite{MandelBook}:
\begin{align}\label{s590}
\exp \bigl( x \hat{A} \bigr) \hat{B} \exp \bigl( -x \hat{A} \bigr) & =  \hat{B} + x \bigl[ \hat{A} , \, \hat{B} \bigr] + \frac{x^2}{2!} \bigl[ \hat{A} ,\bigl[ \hat{A} , \, \hat{B} \bigr]\bigr] + \ldots \;.
\end{align}
To this end, first we evaluate the commutator
\begin{align}\label{s620}
\bigl[\hat{A} , \, \hat{B} \bigr] & = \sum_{n=1}^M \alpha_n \Bigl[ \hat{C}_n , \,  \hat{a}_\nu^\dagger \Bigr] \nonumber \\[6pt]
& =  \sum_{n=1}^M \alpha_n \sum_{\gamma, \beta} \bm{1}_{n \gamma \beta} \underbrace{\Bigl[\hat{a}_\gamma^\dagger \hat{a}_\beta , \,  \hat{a}_\nu^\dagger \Bigr]}_{= \; \delta_{\beta \nu} \hat{a}^\dagger_\gamma} \nonumber \\[6pt]
& =  \sum_{n=1}^M \alpha_n \underbrace{\sum_{\gamma} \bm{1}_{n \gamma \nu}  \hat{a}^\dagger_\gamma}_{\deff \;\hat{\phi}_{n  \nu}^\dagger } \nonumber \\[6pt]
& = \sum_{n=1}^M \alpha_n \hat{\phi}_{n  \nu}^\dagger,
\end{align}
where we have used the relation
\begin{align}\label{s630}
\bigl[\hat{A} \hat{B} , \, \hat{C} \bigr] & =   \hat{A}  \bigl[\hat{B} , \, \hat{C} \bigr]  + \bigl[\hat{A}, \, \hat{C}  \bigr]\hat{B},
\end{align}
with $\hat{A} = \hat{a}_\gamma^\dagger, \, \hat{B} =  \hat{a}_\beta$ and $\hat{C} = \hat{a}_\nu^\dagger$, and we have defined
\begin{align}\label{s635}
\hat{\phi}_{n  \nu}^\dagger \deff \Bigl[ \hat{C}_n , \,  \hat{a}_\nu^\dagger \Bigr] = \sum_{\gamma} \bm{1}_{n \gamma \nu}  \hat{a}^\dagger_\gamma = \bigl( u_\nu, \bm{1}_n \hat{\phi}^\dagger \bigr),
\end{align}
where \eqref{n10} and \eqref{a222} have been used.
The next commutator is
\begin{align}\label{s640}
\bigl[\hat{A} , \,\bigl[\hat{A} , \, \hat{B} \bigr] \bigr] & =   \sum_{n=1}^M \alpha_n \Bigl[ \hat{C}_n , \,  \sum_{m=1}^M \alpha_m \hat{\phi}_{m \nu}^\dagger \Bigr]  \nonumber \\[6pt]
& =     \sum_{n, m=1}^M \alpha_n \alpha_m \Bigl[ \hat{C}_n , \,  \hat{\phi}_{m \nu}^\dagger \Bigr]\nonumber \\[6pt]
& =     \sum_{n, m=1}^M \alpha_n \alpha_m \sum_{\gamma} \bm{1}_{m \gamma \nu}  \underbrace{\Bigl[ \hat{C}_n , \,  \hat{a}^\dagger_\gamma \Bigr]}_{= \; \hat{\phi}_{n \gamma}^\dagger} \nonumber \\[6pt]
& =     \sum_{n, m=1}^M \alpha_n \alpha_m \sum_{\gamma} \bm{1}_{m \gamma \nu}  \hat{\phi}_{n \gamma}^\dagger.
\end{align}
Now we note that from \eqref{s635} we have
\begin{align}\label{s650}
\sum_{\gamma} \bm{1}_{m \gamma \nu}   \hat{\phi}_{n \gamma}^\dagger & =   \sum_{\gamma} \bm{1}_{m \gamma \nu}  \sum_{\tau} \bm{1}_{n \tau \gamma}  \hat{a}^\dagger_\tau  \nonumber \\[6pt]
& =  \sum_{\tau} \left(\sum_{\gamma} \bm{1}_{m \gamma \nu}  \bm{1}_{n \tau \gamma} \right)  \hat{a}^\dagger_\tau  \nonumber \\[6pt]
& =  \delta_{nm} \, \sum_{\tau} \bm{1}_{n \tau \nu}  \hat{a}^\dagger_\tau  \nonumber \\[6pt]
& =  \delta_{nm} \hat{\phi}_{n  \nu}^\dagger ,
\end{align}
where \eqref{s120},\eqref{s150} and \eqref{s200} have been used. Substituting \eqref{s250} into \eqref{s640}, we obtain
\begin{align}\label{s660}
\bigl[\hat{A} , \,\bigl[\hat{A} , \, \hat{B} \bigr] \bigr] & =    \sum_{n=1}^M \alpha_n^2 \hat{\phi}_{n  \nu}^\dagger.
\end{align}
We can iterate the procedure above $k$ times to find
\begin{align}\label{s665}
\bigl[ \underset{\vphantom{\bigl[} k}{\hat{A}} ,  \, \bigl[ \underset{\vphantom{\bigl[} k-1}{\hat{A}} , \, \ldots \bigl[\underset{\vphantom{\bigl[} 2}{\hat{A}} , \, \bigl[\underset{\vphantom{\bigl[} 1}{\hat{A}} , \, \hat{B} \bigr] \bigr]  \ldots \bigr] \bigr] & =     \sum_{n=1}^M \alpha_n^k \hat{\phi}_{n  \nu}^\dagger, \qquad (k=1,2, \ldots, D),
\end{align}
so that
\begin{align}\label{s670}
e^{i \hat{A}} \hat{a}_\nu^\dagger e^{-i \hat{A}}  & =   \hat{a}_\nu^\dagger  + i  \sum_{n=1}^M \alpha_n \hat{\phi}_{n  \nu}^\dagger  + \frac{i^2}{2!}  \sum_{n=1}^M \alpha_n^2 \hat{\phi}_{n  \nu}^\dagger +  \ldots  \nonumber \\[6pt]
& =  \hat{a}_\nu^\dagger  +  \sum_{n=1}^M \hat{\phi}_{n  \nu}^\dagger \left( i \,\alpha_n   + \frac{i^2}{2!} \, \alpha_n^2  +  \ldots \right) \nonumber \\[6pt]
& =  \hat{a}_\nu^\dagger - \sum_{n=1}^M \hat{\phi}_{n  \nu}^\dagger +  \sum_{n=1}^M \hat{\phi}_{n  \nu}^\dagger \left( 1 + i \,\alpha_n   + \frac{i^2}{2!} \, \alpha_n^2  +  \ldots \right) \nonumber \\[6pt]
& =  \hat{a}_\nu^\dagger - \sum_{n=1}^M \hat{\phi}_{n  \nu}^\dagger +  \sum_{n=1}^M \hat{\phi}_{n  \nu}^\dagger \exp \left( i \alpha_n \right).
\end{align}
Substituting \eqref{s670} into \eqref{c10}, we obtain
\begin{align}\label{s680}
\langle 0 | \hat{a}_\mu \exp\left(  i \sum_{n=1}^M \alpha_n \hat{C}_n \right) \hat{a}_\nu^\dagger| 0 \rangle & =   \langle 0 | \hat{a}_\mu \left[ \hat{a}_\nu^\dagger - \sum_{n=1}^M \hat{\phi}_{n  \nu}^\dagger +  \sum_{n=1}^M \hat{\phi}_{n  \nu}^\dagger \exp \left( i \alpha_n \right) \right]   | 0 \rangle \nonumber \\[6pt]
& =  \langle 0 | \hat{a}_\mu  \hat{a}_\nu^\dagger | 0 \rangle  - \sum_{n=1}^M  \langle 0 | \hat{a}_\mu \hat{\phi}_{n  \nu}^\dagger  | 0 \rangle  +  \sum_{n=1}^M e^{ i \alpha_n } \langle 0 |\hat{a}_\mu  \hat{\phi}_{n  \nu}^\dagger     | 0 \rangle  \nonumber \\[6pt]
& =  \delta_{\mu \nu}  - \sum_{n=1}^M  \bm{1}_{n \mu \nu}  +  \sum_{n=1}^M e^{ i \alpha_n } \bm{1}_{n \mu \nu},
\end{align}
where we have used \eqref{s635} to calculate
\begin{align}\label{s685}
\langle 0 |\hat{a}_\mu  \hat{\phi}_{n  \nu}^\dagger     | 0 \rangle & =    \sum_{\gamma} \bm{1}_{n \gamma \nu} \langle 0 |\hat{a}_\mu  \hat{a}^\dagger_\gamma    | 0 \rangle \nonumber \\[6pt]
& =   \sum_{\gamma} \bm{1}_{n \gamma \nu} \delta_{\mu \gamma}   \nonumber \\[6pt]
& = \bm{1}_{n \mu \nu}.
\end{align}
Inserting \eqref{s680} into \eqref{s555}, we obtain
\begin{align}\label{s700}
p_\mathbf{C} (1,\mathbf{c}) & =   \int_{\mathbb{R}} \frac{\mathrm{d}  \alpha_1}{2 \pi} \cdots \int_{\mathbb{R}} \frac{\mathrm{d}  \alpha_M}{2 \pi} \left\{ \exp \left( - i \sum_{n=1}^M \alpha_n c_n \right) \right. \nonumber \\[6pt]
&\phantom{=} \times \left.  \sum_{\mu, \nu} \phi_\mu^* \phi_\nu \left( \delta_{\mu \nu}  - \sum_{m=1}^M  \bm{1}_{m \mu \nu}  +  \sum_{m=1}^M e^{ i \alpha_m } \bm{1}_{m \mu \nu} \right) \right\}.
\end{align}
To evaluate this expression we need to calculate
\begin{align}\label{s710}
\sum_{\mu, \nu} \phi_\mu^* \phi_\nu  \bm{1}_{m \mu \nu} & = \sum_{\mu, \nu} \phi_\mu^* \phi_\nu  (u_\mu, \bm{1}_m u _\nu) \nonumber \\[6pt]
& =  \left( \sum_{\mu} \phi_\mu u_\mu, \bm{1}_m \sum_{\nu} \phi_\nu u _\nu \right) \nonumber \\[6pt]
& =  \left( \phi, \bm{1}_m \phi \right)\nonumber \\[6pt]
& \deff  P_m \geq 0,
\end{align}
where \eqref{s150} has been used. Using this result, we can rewrite \eqref{s700} as,
\begin{align}\label{s720}
p_\mathbf{C} (1,\mathbf{c}) & =   \prod_{n=1}^M \delta \left( c_n \right) \left[\underbrace{\left(\phi, \phi \right)}_{=\; 1}  - \sum_{m=1}^M P_m \right] \nonumber \\[6pt]
& \phantom{=} + \sum_{m=1}^M P_m \int_{\mathbb{R}} \frac{\mathrm{d}  \alpha_m}{2 \pi} \exp \left[ - i (\alpha_m -1) c_m \right] \mathrm{d} \alpha_m \prod_{\substack{n=1 \\[3pt] n \neq m}}^M
\int_{\mathbb{R}} \frac{\mathrm{d}  \alpha_n}{2 \pi} \exp \left( - i  \alpha_n  c_n \right) \mathrm{d} \alpha_n  \nonumber \\[6pt]
& = \prod_{n=1 }^M \delta \left( c_n \right) \left( 1  - \sum_{m=1}^M P_m \right) +
\sum_{m=1}^M P_m \, \delta \left( c_m - 1 \right) \prod_{\substack{n=1 \\[3pt] n \neq m}}^M \delta \left( c_n \right) \nonumber \\[6pt]
& \deff P_0 \, \prod_{n=1 }^M \delta \left( c_n \right) +
\sum_{m=1}^M P_m \, \delta \left( c_m - 1 \right) \prod_{\substack{n=1 \\[3pt] n \neq m}}^M \delta \left( c_n \right),
\end{align}
where \eqref{s250} has been used, and we have defined the probability $P_0$ of zero counting in all photodetectors, as
\begin{align}\label{s730}
P_0 =   1  - \sum_{m=1}^M P_m.
\end{align}
When $M=1$ \eqref{s720} reduces to
\begin{align}\label{s735}
p_C (1,c)  = (1-P) \, \delta \left( c \right) + P \, \delta \left( c - 1 \right) ,
\end{align}
where $P= (\phi, \mathbf{1} \phi)$.
Moreover, note that $p_\mathbf{C} (1,\mathbf{c})$ is correctly normalized because \eqref{s730} trivially implies
\begin{align}\label{s740}
\sum_{m=0}^M P_m =1.
\end{align}
The meaning of this equation is that given the  single-photon state $| 1[\phi] \rangle$, either we get a ``click'' in one of the $M$ detectors, or not.

\section{Probability distribution for the wave-particle operator}\label{pdfWC}

In this appendix we calculate explicitly  $p_{W_1 C_2}(N,w, c)$, defined by
\begin{align}\label{d10}
p_{W_1 C_2}(N,w, c)   = \langle N[\phi] | \delta \bigl( \hat{W}_1 - w \bigr)  \delta \bigl( \hat{C}_2 - c \bigr) | N[\phi] \rangle.
\end{align}

\subsection{Vacuum state}

In this case we have
\begin{align}\label{d20}
p_{W_1 C_2}(0,w, c) & =  \langle 0  | \delta \bigl( \hat{W}_1 - w \bigr)  \delta \bigl( \hat{C}_2 - c \bigr) |  | 0 \rangle \nonumber \\[6pt]
& =  \frac{1}{(2 \pi)^2} \int_{\mathbb{R}} \mathrm{d} \alpha_1  \int_{\mathbb{R}} \mathrm{d} \alpha_2 \, e^{ - i \left( \alpha_1 w +  \alpha_2 c \right) } \langle 0 | e^{i \alpha_1 \hat{W}_1 } \underbrace{ e^{i \alpha_2 \hat{C}_2 }| 0 \rangle}_{= \;| 0 \rangle } \nonumber \\[6pt]
& =  \delta(c) \, \frac{1}{(2 \pi)} \int_{\mathbb{R}} \mathrm{d} \alpha_1  \, e^{ - i  \alpha_1 w  } \langle 0 | e^{i \alpha_1 \hat{W}_1 } | 0 \rangle  \nonumber \\[6pt]
& =  p_{W}(0,w) \, p_{C}(0, c),
\end{align}
because $\hat{C}_2  | 0 \rangle = 0$, and  \eqref{s260} has been used.

\subsection{Single photon state}

In this case, we must evaluate
\begin{align}\label{d30}
p_{W_1 C_2}(1,w, c) & =  \langle 1[\phi] |  \delta \bigl( \hat{W}_1 - w \bigr)  \delta \bigl( \hat{C}_2 - c \bigr)  | 1[\phi] \rangle \nonumber \\[6pt]
& =  \sum_{\mu, \nu} \phi_\mu^* \phi_\nu \langle 0 | \hat{a}_\mu \delta \bigl( \hat{W}_1 - w \bigr)  \delta \bigl( \hat{C}_2 - c \bigr)  \hat{a}_\nu^\dagger | 0 \rangle  ,
\end{align}
where
\begin{align}\label{d40}
\langle 0 | \hat{a}_\mu \delta \bigl( \hat{W}_1 - w \bigr)  \delta \bigl( \hat{C}_2 - c \bigr)  \hat{a}_\nu^\dagger | 0 \rangle  =
\int_{\mathbb{R}} \frac{\mathrm{d}  \alpha_1}{2 \pi} \int_{\mathbb{R}} \frac{\mathrm{d}  \alpha_2}{2 \pi} \, e^{ - i \left( \alpha_1 w +  \alpha_2 c \right) }  \langle 0 | \hat{a}_\mu e^{i \alpha_1 \hat{W}_1 }  e^{i \alpha_2 \hat{C}_2} \hat{a}_\nu^\dagger| 0 \rangle.
\end{align}
We need to calculate explicitly the last term of the previous equation, that is
\begin{align}\label{d50}
\langle 0 | \hat{a}_\mu e^{i \alpha_1 \hat{W}_1 }  e^{i \alpha_2 \hat{C}_2} \hat{a}_\nu^\dagger| 0 \rangle & = \langle 0 | \hat{a}_\mu e^{i \alpha_1 \hat{W}_1 } \left( e^{i \alpha_2 \hat{C}_2} \hat{a}_\nu^\dagger e^{-i \alpha_2 \hat{C}_2} \right) | 0 \rangle \nonumber \\[6pt]
& = \langle 0 | \hat{a}_\mu \, e^{i \alpha_1 \hat{W}_1 } \hat{a}_\nu^\dagger | 0  \rangle - \left( 1-e^{i \alpha_2}\right) \langle 0 | \hat{a}_\mu \, e^{i \alpha_1 \hat{W}_1 } \hat{\phi}_{2 \mu}^\dagger | 0  \rangle  ,
\end{align}
where all the quantities are defined as in the previous two appendices.
The rest of the calculation is very straightforward and yields:
\begin{align}\label{d60}
p _{W_1 C_2}  (1,w,c)   = p_{W}(1,w) \, p_{C}(0,c) + p_{W}(0,w) \, p_{C}(1,c)   - p_{W}(0,w) \, p_{C}(0,c).
\end{align}
Using \eqref{s550}, \eqref{s735}, \eqref{s365} and \eqref{q610}-\eqref{q620}, we can rewrite this equation as
\begin{align}\label{d70}
p _{W_1 C_2}  (1,w,c)  & = \bigl( 1 - |s|^2 - P \bigr) \, p_0(w) \, \delta(c) + P \, p_0(w) \, \delta(c-1) + |s|^2 \, p_1(w) \, \delta(c) \nonumber \\[6pt]
& \deff \bigl( 1 - |s|^2 - P \bigr) \, p_{00}(w,c) \, \delta(c) + P \, p_{01}(w,c) + |s|^2 \, p_{10}(w,c) ,
\end{align}
where we have defined
\begin{align}\label{d80}
p_{00}(w,c) \deff p_0(w) \, \delta(c), \qquad   p_{01}(w,c) \deff p_0(w) \, \delta(c-1), \qquad \text{and} \qquad p_{10}(w,c)  \deff p_1(w) \, \delta(c) .
\end{align}

\end{document}